\documentclass{aa}  

\usepackage{graphicx}
\graphicspath{ {./images/} }
\usepackage{subcaption}
\usepackage{txfonts}
\usepackage{pdfpages}
\usepackage{multirow}

\usepackage{natbib}
\bibpunct{(}{)}{;}{a}{}{,}
\usepackage{hyperref}

\begin{document}

   \title{Spiral arms across stellar populations in simulations \\ via the local dimension method}

   \author{
        J. Ard\`evol \inst{1,2,3}   \fnmsep
        \thanks{\email{jardevol@icc.ub.edu}}
        \and
        M. Semczuk \inst{1,2,3}
        \and
        T. Antoja \inst{1,2,3}
        \and
        V.P. Debattista \inst{4}
        \and
        M. Bernet \inst{1,2,3}
        \and
        A. Pettitt \inst{5}
           }
   \institute{
        Institut de Ci\`encies del Cosmos (ICCUB), Universitat de Barcelona (UB), c. Mart\'i i Franqu\`es 1, 08028 Barcelona, Spain
        \and
        Departament de Física Qu\`antica i Astrof\'isica (FQA), Universitat de Barcelona (UB), c. Mart\'i i Franqu\`es 1, 08028 Barcelona, Spain
        \and
        Institut d'Estudis Espacials de Catalunya (IEEC), Edifici RDIT, Campus UPC, 08860 Castelldefels (Barcelona), Spain
        \and
        Jeremiah Horrocks Institute, University of Central Lancashire, Preston, PR1 2HE, UK
        \and
        Department of Physics and Astronomy, California State University, Sacramento, 6000 J Street, Sacramento, CA 95819, USA
              }

   \date{Received XXX; Accepted YYY}

    \abstract
    { The origin and nature of spiral arms remain unclear. Star forming regions and young stars are generally strongly associated to the spiral structure, but there are few quantitative predictions from simulations about the involvement of stars of different ages.}
    { We aim to quantify the interplay between spiral arms and different populations.}
    { We use a hydrodynamical simulation of an isolated disc galaxy displaying a dynamic multi-armed spiral structure. Inspired by cosmological structure metrics, we develop a new method, the local dimension, that robustly delineates arms across populations and through space and time.}
    { We find that all stars, including those as old as 11\,Gyr, support the arms. The spiral strength decreases with stellar age up to 2\,Gyr-old stars and remains nearly constant for older stars. However, the scaling between arm strength and age (or velocity dispersion) depends on the strength of the global spiral structure at each time. Almost all stars formed in arms remain within them for no more than 140--180\,Myr, whereas old stars leave arms about three times faster. Even if the youngest populations dominate in the production of the spiral torques at early times, all populations contribute equally at later times.}
    {Our results highlight the power of the local dimension for studying complex spiral structures and show that all stellar populations in the disc partake in the arms. Since in our model we see spiral arms in populations with velocity dispersions up to 90\,km/s, which are comparable to those of the Milky Way, we predict that old Galactic populations could also exhibit spiral structure.
    }

   
   \titlerunning{Spiral arms across populations}
   \authorrunning{J. Ard\`evol et al.}
   
   \maketitle

%-------------------------------------------------------------------
\section{Introduction}

    Spiral galaxies constitute about 60\% of the local galaxies \cite[e.g.][]{Nair10_fractSpiralGal,Lintott11_fractSpiralGal,Willett13_fractSpiralGal}.
    Yet, despite decades of study, certain aspects of spiral structure remain actively debated: which are the spiral excitation mechanisms and dynamics, and are they long-lived density waves or transient, recurrent features \citep[reviews by][and references therein]{DobbsBaba14, SellwoodMasters22}.

    Observations of external galaxies reveal the connection between spiral arms and star formation, with young stellar populations, and molecular and ionized gas tracing the arms \citep{Rand95_M100COarms,Seigar02_IRarmsEnhanceSF,Pessa23_extragalArmsYounger,Garner24_MultiplePatterns}. Old stellar populations, such as those traced in red or infrared (IR) light, also delineate spiral structure, although the resulting structures are often smoother than for young stars \cite[e.g.][]{Zwicky55_ExtragalArmsAllPops,Schwizer76_ExtragalArmsAllPops,Elmegreen81_OptVSIR,Rix95_Karms,Jarrett03_IRdecoupledOpt}.
    Some galaxies which appear flocculent or multi-armed in optical light have underlying two-armed grand-design spiral structures in near-IR \citep{Block94_youngVSoldArms, Thronley96_OptFlocIRgd, Kendall11_externalOldArms}.
    In contrast, other authors have found that both IR and optical observations lead to consistent classifications irrespective of the galaxy type \citep{Eskridge02_OptSimilarMIR,Buta10_IRbluemorphology,Elmegreen11_IRarms,Kendall11_externalOldArms}.

    For the Milky Way, population-dependent detections of spiral arms are further complicated by our position within the disc, the distance uncertainties and the effects of dust obscuration. These hinder a clear view of the spiral structure and may lead to inconsistencies in the precise location of the spiral arms and their exact number among different tracers.
    Our Galaxy's spiral arms were originally discovered through the alignment of nearby star forming regions \citep{Morga1952_1stMWarmsHIIOB}, and were confirmed using HI maps \citep{Oort1958_1stMWarmsHI,Levine2006_MWarmsHI, Nakanishi2016_gasMWarms, Koo2017_MWarmsHI}, which covers large Galactic scales thanks to the low extinction at radio wavelengths.
    \cite{Reid2019_MWspiralArms} studied the maser emission of extremely young massive stars and found four major arms, plus a local segment.
    \cite{Hou21_MWspiralArms} re-derived the parameters of these five arms by combining several young tracers (molecular clouds, masers, HII regions, O-type stars and young open clusters).
    In contrast, \cite{Xu23_MWmultiarmed} used similar targets to conclude that the Milky Way may have two inner spiral arms and a multi-armed structure at large radii.
    Local samples of young stars generally show a complex structure with clumps rather than continuous overdensities \citep{Poggio2021_MWarmsUMS, Zari21_MWarmsOBA, Ardevol23_MWarmsA}.
    On the other hand, \cite{Lin22_MWarmsRC} and \cite{Uppal23_MWarmsRC} detected the Outer arm, and the Local and Perseus arms, respectively, using old red clump stars. Indeed, old stars covering more extended regions seem to present two major arms \citep{Drimmel00_2oldMWarms,Drimmel01_KvsOpticalMWarms,Benjamin09_2MWmidIRarms,Khanna24_oldMWarms}.
    Differences among populations may be physical as in some external galaxies, or due to the aforementioned observational effects.
    
    Recently, it has become possible to measure global quantities, such as velocity dispersion, age and metallicity, on top of the spiral arms in external galaxies.
    \cite{Breda24_externalArmsAVR} found lower mean age and lower velocity dispersion along the spiral arms in NGC 4030 with data from the Multi Unit Spectroscopic Explorer \cite[MUSE,][]{Bacon10_MUSE}. \cite{Chen24_MAGPI2gal} used data from the same instrument and concluded that the amplitude of metallicity and age azimuthal asymmetries depends on pitch angles through the mixing timescale.
    \cite{SanchezMenguiano20_ExtArmsChem} studied the enhancements in the gas metallicity of external galaxies and found that they correlate with spiral arms in about half of the cases.
    In the Milky Way, it has been reported that spiral arm segments tend to exhibit slightly enhanced metallicities --by around 0.1\,dex \citep{Poggio22_ArmsChemistry,Hawkins23_ArmsChemistry,Hackshaw24_ArmsChemistry,Barbillon25_ArmsChemistry,ViscasillasVazquez25_ArmsChemistry}. 
    These trends may reflect the presence of recently formed stars, which are typically more metal-rich and kinematically colder, or alternatively,     may result from stars in dynamically cold orbits being preferentially concentrated in the arms (as suggested by \citealt{Khoperskov18_ArmsFeHkinPops}, \citealt{Palicio23_MWarmsJR} or \citealt{Debattista24_sim}).
    Understanding the contribution of different populations to the spiral arms is increasingly important in view of all these recent observations and the wealth of data becoming available from ongoing integral field unit surveys \citep{Sanchez12_ifuCALIFA,ErrozFerrer19_ifuMUSE,Jin24_ifuWEAVE} and modern spectroscopic studies of Milky Way stars.
    
    Hydrodynamical simulations of galaxy evolution that incorporate star formation are essential for understanding spiral structure and the observed complexity across different stellar populations. As discussed in \citet{Sellwood84_gasneeded}, the role of gas and new generations of young stars on cold orbits is crucial to maintain the spiral structure in the disc, which otherwise would heat up, stopping the formation of arms (as in e.g. \citealt{Bird13_SimPops}).
    \citet{Grand12_noOffset} studied a hydrodynamical simulation of an isolated barred galaxy within a static dark matter halo and found some broadening of the arms with age. \citet{Pettitt17_starFormation} compared the distributions of gas, young and old stars in their $N$-body models with star formation and found that spiral arms are present in all three tracers. \citet{Khoperskov18_ArmsFeHkinPops} concluded that dynamically cold stars support stronger spiral arms but this was done using an $N$-body simulation of three kinematically distinct populations (cool, warm and hot) instead of including gas and star formation.
    \cite{Ghosh22_oldSimArms} found spiral arms for all stars in their isolated high-resolution hydrodynamical simulation, which is the one that we study here, although they focused on the vertical breathing modes associated with the arms.
    \citet{Debattista24_sim} analysed this same model and found that younger stellar populations, being dynamically cooler, support the spiral structure more effectively, which leads to metal-rich signatures coincident with the spiral arms.
    \cite{Palicio25_oldSimArms} found spiral arms in particles as old as 3\,Gyr in the Auriga \citep{Grand17_Auriga,Grand24_Auriga} and IllustrisTNG \citep{Nelson24_Illustris2,Nelson19_Illustris,Pillepich19_Illustris2,Pillepich24_IllustrisMWlike} cosmological simulations, but concluded that the g2.79e12 NIHAO-UHD simulation \citep{Buck20_NIHAO} seem to have spiral structure only in their youngest component.
    
    Taking advantage of the latest generation of high-resolution simulations with more realistic physics, we aim to improve and expand these limited existing predictions by providing a detailed characterization of the spiral arms across stellar populations. Specifically, we explore the isolated galaxy disc of the state-of-the-art hydrodynamical simulation used by \cite{Debattista24_sim}.
    By tracing both gas and stars of different ages, we examine how spiral arms morphology, strength and dynamical impact vary across populations. For this, we introduce a novel method to detect and determine the properties of spiral arms based on the geometry of the density field, quantified through the local dimension, $D$.
        
    We describe the simulation setup and our methods in Sects.~\ref{sectData} and \ref{sectMethod}, respectively. In Sect.~\ref{sectMorpho} we explore the general characteristics and morphology of the spiral structure for different stellar ages and in Sect.~\ref{sectEvolution} we perform a detailed analysis of the strength of the arms across different populations. To explain the age dependence of the spiral strength, in Sect.~\ref{sectTrace} we address the persistence within spiral arms of several populations. In Sect.~\ref{sectTorques} we quantify how important each population is to the non-axisymmetric forces in the disc through the torques produced by each of them. Section~\ref{sectDiscussion} discusses our results and compares them with the previous literature. We summarise our conclusions in Sect.~\ref{sectConclusions}.

%--------------------------------------------------------------------
\section{Simulation}\label{sectData}

    To study spiral arms across stellar populations, we used the model M1\_c\_b from the $N$-body+smooth particle hydrodynamics simulations suite presented in \cite{Fiteni21_sim}.
    This model was previously used by \cite{Khachaturyants22_simwarp,Khachaturyants22_simvert} and \cite{Ghosh22_oldSimArms} to study the vertical bending and breathing motions in the disc, and by \cite{Debattista24_sim} to analyse azimuthal metallicity variations related to spiral arms. Recently, \citet{Bernet25_DarkSpirals} studied the interaction between the spiral arms and the dark matter halo in a suite of simulations, which included this model.
    
    The simulation follows an isolated spiral galaxy with an ever-changing multi-armed spiral structure. 
    The galaxy is embedded in a live dark matter halo with virial mass $M_{200}$\,=\,$10^{12}\,M_{\odot}$ and virial radius $R_{200}$\,$\sim$\,200\,kpc, represented by 5 million particles with masses of $2\times10^{5}\,M_{\odot}$ inside $R_{200}$ and $7.1\times10^{5}\,M_{\odot}$ beyond. The baryonic component starts as a hot gas halo with a mass of about $10^{11}\,M_{\odot}$ represented by 5 million gas particles with an initial mass of $2.8\times10^4$\,$M_{\odot}$ each. Gas from this corona cools down to form a disc, which undergoes a short episode of clumpy star formation. Stellar particles, each of a mass of $9.3\times10^3$\,$M_{\odot}$, are formed from gas particles that have a temperature lower than $1.5\times10^{4}$\,K and a density exceeding 0.1\,$\mathrm{amu\,cm}^{-3}$, and that are in a convergent flow.
    The model uses a gravitational softening parameter of $\epsilon$\,=\,50\,pc for both gaseous and stellar particles (from now on called gas and stars, respectively).
    The simulation ends after 13\,Gyr, resulting in a total of 11\,587\,120 stars. At these later stages of the run, the gaseous and stellar discs have masses of the order of $10^{9}\,M_{\odot}$ and $10^{10}\,M_{\odot}$, respectively.
    These masses, the baryonic fraction and the radial scale lengths of all components are comparable to the Milky Way values \citep{BlandHawthorn17_MWprop, Debattista24_sim}.

    The disc of this simulated galaxy forms a relatively weak bar at about 4\,Gyr that weakens at around 7\,Gyr, reforms shortly later, and then weakens considerably around 11--13\,Gyr (Fig.~2 from \citealt{Fiteni21_sim} and Fig.~7 from \citealt{Debattista24_sim}).
    To focus solely on spiral arms and to avoid regions that are largely affected by the dynamics of the bar and the nuclear stellar disc, we exclude the inner 4\,kpc in all our analysis. Nevertheless, some overlap between the outer Lindblad resonance of the bar (around 6\,kpc) and spiral arms analysed in our region of interest may still be present. We discuss later the possible effect of the bar on the spiral arms; however, since it is weak and later dissolves, we do not expect it to drastically affect the conclusion of our study.
    In this work we analyse snapshots every 500\,Myr from 5 to 13\,Gyr, and for parts of our analysis we make use of the higher cadence of 5\,Myr that is available.

%--------------------------------------------------------------------
\section{Methods}\label{sectMethod}

    \begin{figure*}
    \begin{center}
        \includegraphics[width=0.8\textwidth]{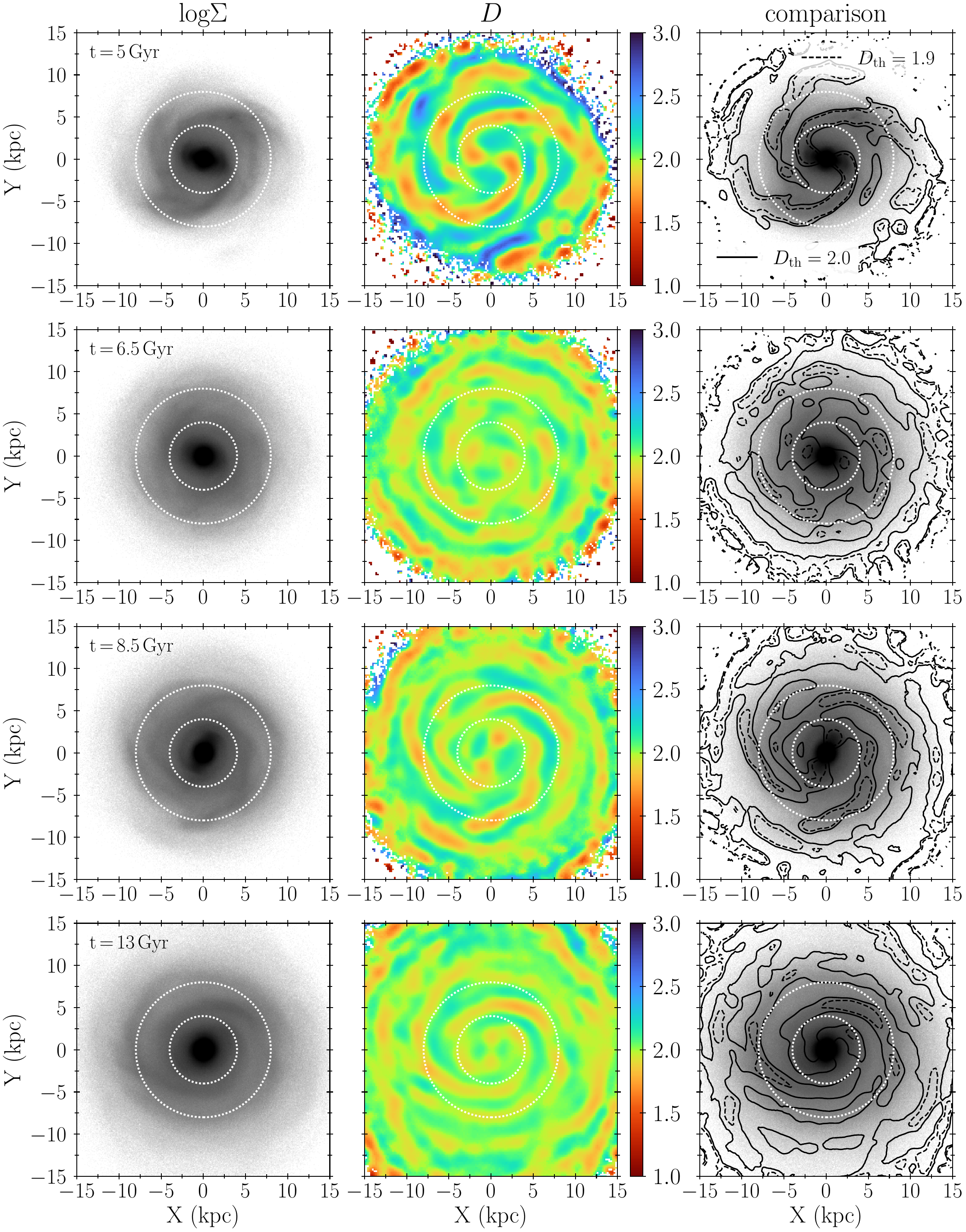}
        \caption{Comparison between the local dimension, $D$, and the density, $\mathrm{log\Sigma}$, in the model, considering all stars. Rows show the times 5, 6.5, 8.5 and 13\,Gyr from top to bottom. {\it Left}: surface density map; {\it middle}: local dimension map; {\it right}: the same as the left panels but with contours at $D=1.9$ (dashed) and $D=2$ (solid) overplotted. White dotted circles show radii of 4 and 8\,kpc, which are the limits used in most of our analysis. Rotation is anticlockwise. }
        \label{fig_DandDensitytime}
    \end{center}
    \end{figure*}
    
    Traditional techniques to detect spiral arms usually rely on Fourier expansions (based on discrete symmetric modes, e.g. \citealt{Pettitt16_gasStarArms}), kernel functions \cite[e.g.][]{Poggio2021_MWarmsUMS}, logarithmic spiral fittings \citep[e.g.][]{SellwoodEA1986,Reid2019_MWspiralArms,Querejeta24_contrasts}, assumptions about the galactic density profile \citep[e.g.][]{Chen24_MAGPI2gal,Kalita25_z1arms} or by-eye tracing \citep[e.g.][]{Masters21_GalaxyZooArms,Palicio25_oldSimArms}, among others. Here we present a new flexible yet objective method that we use to characterize the spiral arms in our simulation. This allows us to unambiguously delineate the arms location without the need for user guidance or shape assumptions.

    The method is inspired by the local dimension used in cosmology to describe the geometry of large-scale galactic environments \citep{SarkarBharadwaj08_D, Sarkar19_D,Das23_D}.
    To measure the local dimension at a given position in the original formalism, we would count particles inside concentric spheres with increasing radius, centred at that position.
    The number of particles inside each sphere, $N$, depends on the radius of the sphere, $r$, and can be locally approximated by 
    \begin{equation}\label{eqD}
        N(<r) = \alpha r^D,
    \end{equation}
    where $D$ is the local dimension of the studied point and $\alpha$ is a constant related to the density normalization \citep{SarkarBharadwaj08_D,Das23_D}.
    We use a similar approach to distinguish between the underlying two-dimensional galactic disc and the spiral arms that can be approximated as filament-like structures ($D$\,<\,2). Interarm regions will present local dimensions larger than two since density around them increases faster than in a homogeneous two-dimensional distribution due to the neighbouring arms.
    
    To simplify the computation of the local dimension for studying disc substructures, we used cylinders of infinite height along the vertical coordinate instead of spheres.
    For our essentially flat discs, this simplification yields very similar results compared to the 3D case and is more similar to what can be done observationally in external galaxies.
    The values of $D$ and $\alpha$ are estimated through a linear regression of Eq.~\ref{eqD} in some radial range $[r_{\mathrm{min}}, r_{\mathrm{max}}]$ (we present examples of fits in Appendix~\ref{app_Dfits}). This range needs to be adapted to the scale of the structures being studied.
    Inappropriate boundaries can blur neighbouring features or highlight irrelevant small-scale features. In this study, we take $r_{\mathrm{min}}$\,=\,0.2\,kpc and $r_{\mathrm{max}}$\,=\,2\,kpc since arms widths are of the order of 1\,kpc \citep{Honig15_armswidth,SilvaVilla22_armswidth,Savchenko20_armswidth}.
        
    While a perfectly uniform two-dimensional distribution would have $D$\,$\sim$\,2 everywhere (except at the edges), an exponential disc would still have predominantly $D$\,$\sim$\,2 in all its positions but would yield lower values of $D$ in the inner parts due to the density gradient \footnote{Effectively, the centre is point-like, with local dimension $D$\,$\sim$\,0.}.
    To remove the effect of this gradient in our calculations, we weighted each particle with the inverse of the surface density at its radius.
    We calculated this through the azimuthally averaged radial density profile of the disc from the \texttt{Profile} class from \textsc{pynbody}\footnote{\url{https://github.com/pynbody/pynbody}}.
    With this correction, any axisymmetric disc will have $D$\,$\sim$\,2 everywhere, with small deviations due to random noise near the outskirts.
    
    We computed $D$ at each point of a Cartesian grid of 0.2\,kpc spacing on X and Y with the galaxy angular momentum aligned with the Z axis. Then, we imposed a resolution cut by keeping $D$ only where there is at least 1 particle inside $r_{\mathrm{min}}$ (i.e. $N(<r_{\mathrm{min}})\equiv N_{\mathrm{min}}\geq1$). 
    Those grid points where $N_{\mathrm{min}}$\,=\,1 have on average around 170 particles inside $r_{\mathrm{max}}$ throughout the evolution and are always restricted to the disc outskirts. This cut minimizes the effects of artifacts at the outer parts of the disc where the number of particles is extremely low.
    
    Figure~\ref{fig_DandDensitytime} presents the spiral structure of our model at different times of the simulation.
    The first column shows the surface density distributions of all stars, while the second one presents the corresponding $D$ maps. The right column also presents the surface densities, but with superimposed contours corresponding to $D$\,=\,1.9 and $D$\,=\,2. These contours clearly delimit regions of the highest density, corresponding to spiral arms. At early times (e.g. 5\,Gyr, first row) the arms are quite strong and exhibit low $D$ values (intense red colours). A clear two-armed structure is seen together with a third arm and hints of a fourth one. At 6.5\,Gyr (second row), the spiral arms are weaker, as evidenced by $D$ values more concentrated around 2 and less defined structures. At later times (8.5 and 13\,Gyr, last two rows), the arms are again better defined, which is overall in agreement with the measured power in the Fourier modes at different times in Fig.~7 of \cite{Debattista24_sim}. In Fig.~\ref{fig_DandDensitytime}, we can see how the $D$ maps highlight weak segments of arms that are not so well seen in the surface density. Besides, thanks to the radial weighting described above, we are able to emphasise the bar-to-arms transition, and to see the arms towards outer radii where the overall density is low.
    
    \begin{figure}
    \begin{center}
        \includegraphics[width=0.42\textwidth]{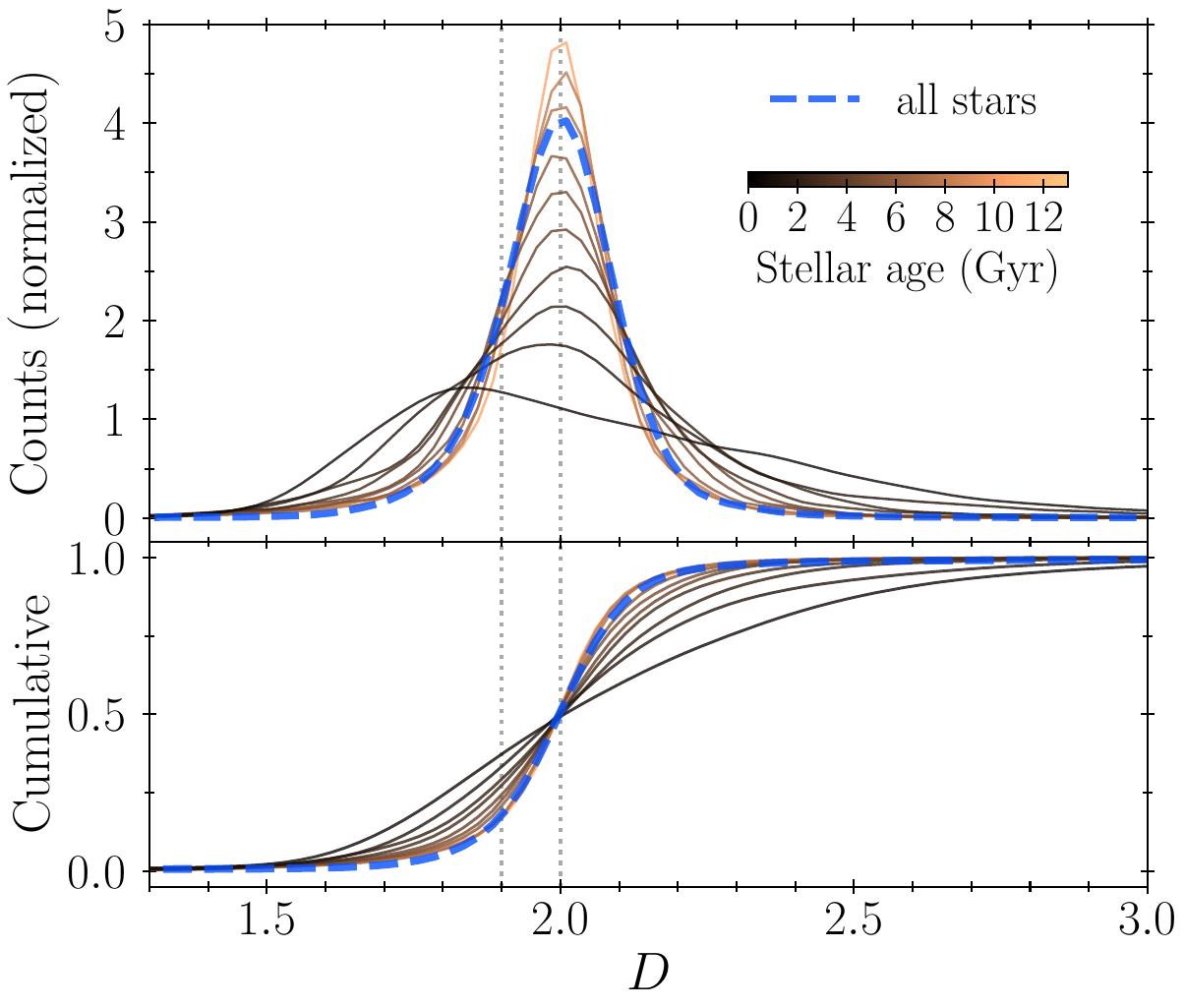}
        \caption{Normalised (top) and cumulative distributions (bottom) of local dimension values from all stars (blue dashed curve) and from different stellar populations (solid curves) at 13\,Gyr. Grid points at radius smaller than 4\,kpc are excluded. Vertical grey dotted lines indicate the reference values $D$\,=\,1.9 and $D$\,=\,2.}
        \label{fig_Dkde}
    \end{center}
    \end{figure}

    For some of our analysis, it will be useful to define the exact location and shape of the spiral arms. For this, we choose a certain threshold of the local dimension, typically $D_\mathrm{th}$\,$\lesssim$\,2. Spiral arms will be taken as regions with $D$\,<\,$D_\mathrm{th}$, since these would correspond to structures of lower dimensionality. We define the boundary of the spiral arms as contours where $D$\,=\,1.9 ($D_\mathrm{th}$\,=\,1.9) for stars younger than 0.5\,Gyr. We take the young population as reference since it usually presents better defined spiral arms, as we will see later on in detail. The exact value of $D_\mathrm{th}$ is chosen following the discussion in Appendix~\ref{app_Dfits}. Our results and conclusions do not change if this arbitrary threshold is changed within a reasonable range.
    While this is the default definition used in our work, the four parameters $r_{\mathrm{min}}$, $r_{\mathrm{max}}$, $N_{\mathrm{min}}$ and $D_{\mathrm{th}}$ can be adjusted to extract different information depending on the context.
    
    In our analysis we use two different measures of the strength of the spiral arms, for instance to compare quantitatively different populations. The first one is the standard deviation of the $D$ values, $\sigma_D$, which is a geometrical definition that indicates the amount of substructure. Figure~\ref{fig_Dkde} shows the distribution of the values of $D$ accross the whole disc excluding the inner part at the time 13\,Gyr computed using all stars (blue dashed line) and as a function of stellar age (thinner solid lines).
    In all cases, the vast majority of grid points have values between $D$\,=\,1.5 and $D$\,=\,2.5. However, older populations present clearly narrower distributions (smaller $\sigma_D$) than younger ones.
    This is because young stars show large density variations: they have strong arms (with low local dimension) separated by scarcely populated interarms (with high $D$). Old stars, instead, are more uniformly distributed across the disc.  Therefore, a high dispersion indicates significant spatial heterogeneity or irregularity in the density distribution, related to the spiral arms and the interarm regions.

    The other measure of the arm strength is the more commonly used arms-vs-disc density contrast of the spiral arms. For this, we take the outlines of the arms (contour of $D_{\mathrm{th}}$\,=\,1.9 for stars younger than 0.5\,Gyr, as explained above) in the radial range 4--8\,kpc, we measure the mean surface density within these spiral arms, $\Sigma_{a}$, and the mean surface density of the disc within the same ring, $\overline{\Sigma}$, and compute the density contrast as:
    \begin{equation}\label{sigma}
    \delta_{\Sigma}=\frac{\Sigma_{a}-\overline{\Sigma}}{\overline{\Sigma}}.
    \end{equation} 
    Positive values of $\delta_{\Sigma}$ indicate that the spiral arms are denser than the azimuthally averaged density in the disc. For instance, $\delta_{\Sigma}$\,=\,0.2 means that the arms are 20\% denser than the disc average at those radii (i.e. $\Sigma_{a}$\,=\,$1.2\overline{\Sigma}$). In a similar way, one can easily compute the arms-vs-interarms density contrast, $\Sigma_{a}/\Sigma_{i}$, where $\Sigma_{i}$ is the mean surface density in the interarms, defined as regions with $D$\,>\,$D_{\mathrm{th}}$.

    Our approach based on $D$ presents a new, model-independent methodology for spiral arms detection.
    The local dimension, defined purely in geometric terms, is versatile in its ability to identify spiral structures with any number of arms, including those with complex or irregular shapes.
    Notably, even arms with lower amplitude are effectively highlighted.
    This methodology provides a precise definition of the arms locations and facilitates detailed comparisons of structural changes across different times, stellar populations, simulations or even observational datasets.

%--------------------------------------------------------------------
\section{Results}\label{sectResults}
    
    The method described above easily outlines the location of spiral arms and helps in quantifying the arms properties.
    Below we consider how spiral arms look like in different populations as outlined by $D$ and how the arm strength varies among populations, and with time. We study how stars of different ages are associated with the arms delineated by our method and how they contribute to the non-axisymmetric forces in the disc.

% ------------------------------------------
\subsection{Spiral structure and morphology as defined by the local dimension across populations }\label{sectMorpho}

    \begin{figure*}
    \begin{center}
        \includegraphics[width=0.9\textwidth]{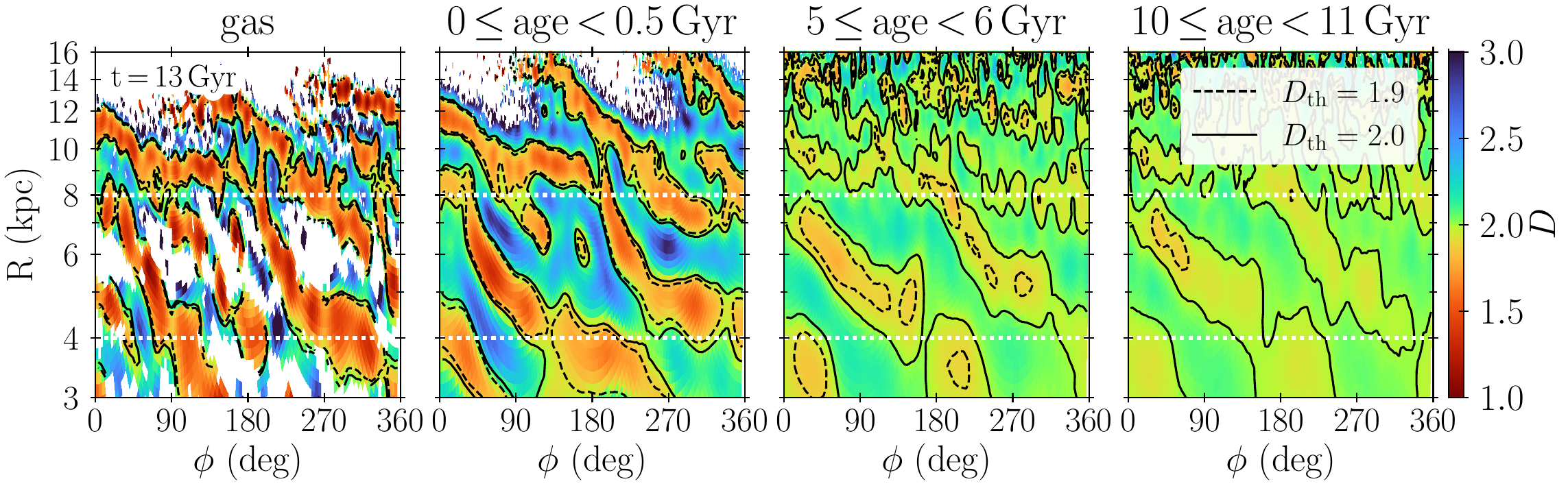}
        \caption{Local dimension maps in log-polar coordinates for gas (leftmost panel) and stars of different ages (younger than 0.5\,Gyr, between 5 and 6\,Gyr-old, and between 10 and 11\,Gyr-old, from left to right) at time 13\,Gyr. The azimuth increases in the direction of rotation. Contours of $D$\,=\,1.9 (dashed) and $D$\,=\,2 (solid) are overplotted. Radii 4 and 8\,kpc are indicated with white dotted lines.}
        \label{fig_DandDensity}
   \end{center}
   \end{figure*}
   
    Figure~\ref{fig_DandDensity} presents the spiral structure of our model at the end of the simulation, at 13\,Gyr, in log-polar coordinates for gas\footnote{We define the gaseous disc as low temperature gas particles with $T$\,$\leq$\,$1.5\times10^{4}$\,K (the temperature required to form stars) and having vertical galactocentric coordinate satisfying $|z|$\,$\leq$\,2\,kpc. By doing this, we avoid the hot gas corona.} and different stellar-age selections: stars younger than 0.5\,Gyr, stars with ages between 5 and 6\,Gyr, and old stars between 10 and 11\,Gyr-old. Holes in the gas panel are due to interarms being almost devoid of cold gas ($N_{min}$\,=\,0). This figure shows that even the oldest stellar populations exhibit spiral structure, albeit of lesser intensity than young stars.
    Globally, we see regions with both lower and larger values of $D$ for the youngest sample (larger $\sigma_D$) than for the older ones, indicative of stronger spiral arms.
    We quantify this in more detail in Sect.~\ref{sectEvolution} while here we discuss the global morphology of the arms across populations. 
  
    In Fig.~\ref{fig_DandDensity} we see that arms at different ages are, overall, quite similar in the range going from
    $\sim$\,4\,kpc to $\sim$\,8--10\,kpc, especially when the spiral structure is stronger. The arms manifest as linear features in $\mathrm{log}(R)$--$\phi$ space for the old stars but slightly curved for the young ones. Thus young arms are not exactly logarithmic in this model at this time.
    Beyond $\sim$\,8--10\,kpc, old stars have a less clear spiral structure than young ones.
    Older populations always have larger velocity dispersion than young ones in the simulation. This implies that old stars are less responsive than young ones to non-axisymmetric features, such as spiral arms \citep{Toomre64_Qstability}.
    This responsiveness to perturbations also determines the radial extension of the spiral arms, in the sense that kinematically hotter populations (old stars) present less-extended arms \citep{Athanassoula84}.
    
    We also see that the spiral structure is mainly two-armed for old stars, while gas and young populations show a multi-armed pattern with more small-scale substructure. This is not surprising since the number of spiral arms is also associated with how different populations create and react to the spiral perturbation \citep{Donghia13_SwingAmpl,Michikoshi16_NumberArms}.

% ------------------------------------------
\subsection{Spiral arm strength in different populations}\label{sectEvolution}

 In the previous section, we saw that spiral structure is present in all stellar populations, albeit with younger stars having the strongest signal (Fig.~\ref{fig_DandDensity}). In this section, we quantify this aspect using $\delta_\Sigma$ and $\sigma_D$.
    
    \begin{figure}
        \includegraphics[width=0.5\textwidth]{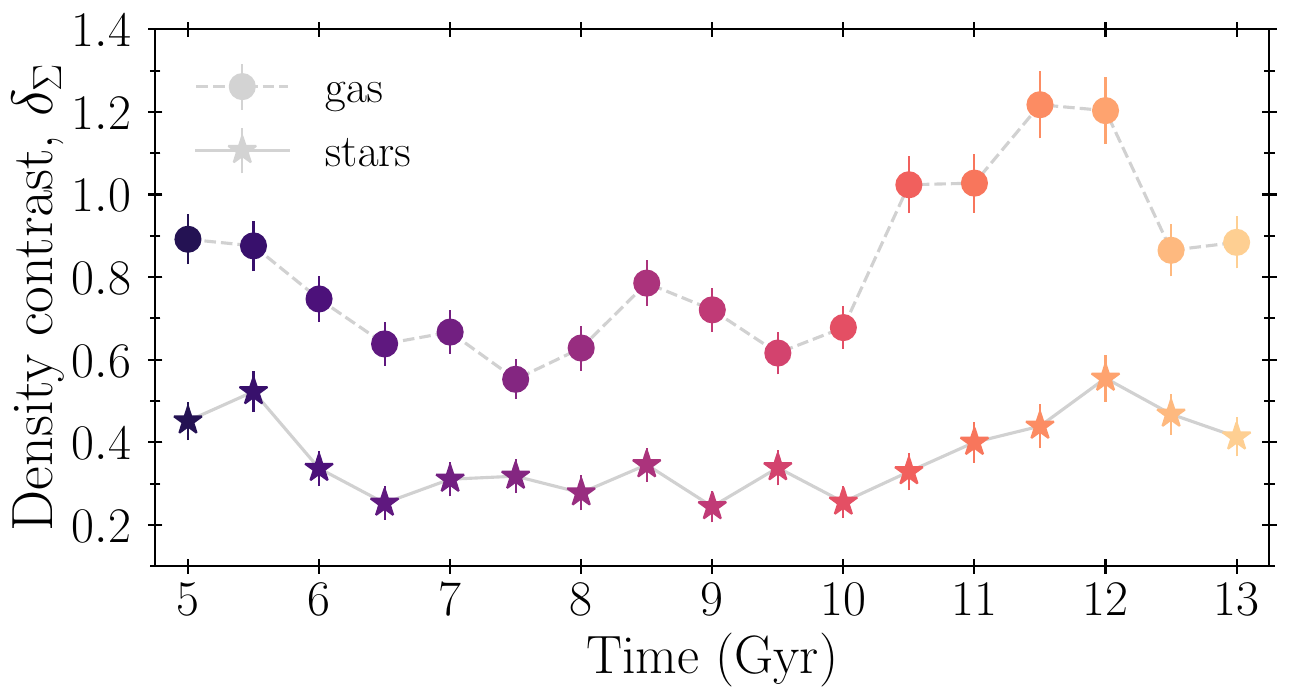}
        \caption{Time evolution of the spiral arms density contrast. The points show the relative ratio between the mean mass surface density of the arms and the disc average mass surface density of gas (circles) and of all stellar populations together (stars) as a function of time. Only galactocentric radii between 4 and 8\,kpc are considered. Colour changes with time as in Figs.~\ref{fig_agehist}, \ref{fig_veldisp} and \ref{fig_torques}.}
        \label{fig_surfdenscontrasts}
    \end{figure}

    In Fig.~\ref{fig_surfdenscontrasts} we show the temporal evolution of the global density contrast of gaseous (circles) and stellar (stars) spiral arms relative to the average disc density. We measured $\delta_{\Sigma}$ as defined in Eq.~\ref{sigma} within the 4--8\,kpc radial ring but the conclusions do not change when the location of the edges of this ring are slightly modified.
    The stellar spiral arms are strong at early times and at late times, with overdensities above 40\% that of the average disc, and they remain moderate between 6 and 10\,Gyr ($\delta_{\Sigma}$\,$\sim$\,30\%). This evolution with time is consistent with the spiral structure depicted by $D$ in Fig.~\ref{fig_DandDensitytime} and with the findings of \citet{Debattista24_sim} (their Fig.~7), where the amplitude of the dominant spiral modes are high at early times and increase after 10.5\,Gyr.
    We notice that gaseous spiral arms have, on average, density contrasts that are twice as large as stellar arms and values that indicate that a large fraction of the gas is actually in spiral arms ($\delta_{\Sigma}$ between 60\% and up to 120\%, i.e. $\Sigma_{a}$\,=\,1.6--2.2\,$\overline{\Sigma}$).
    
    We also computed arm-vs-interarm density contrasts $\Sigma_{a}/\Sigma_{i}$ for all stars and for gas, obtaining values in the ranges 1.4--2.1 and 2.3--3.7, respectively, as shown in Fig.~\ref{fig_contrastsApp} in Appendix~\ref{app_contrasts}.
    Those values show a temporal evolution similar to that of $\delta_{\Sigma}$ in Fig.~\ref{fig_surfdenscontrasts} (i.e. stronger arms both at early and late times).
    Our arms-vs-interarms contrasts approximately agree with the ranges found by \cite{Querejeta24_contrasts} for non-grand-design galaxies, although our values tend to lie toward the upper end of their distribution.
    
    \begin{figure}
        \includegraphics[width=0.48\textwidth]{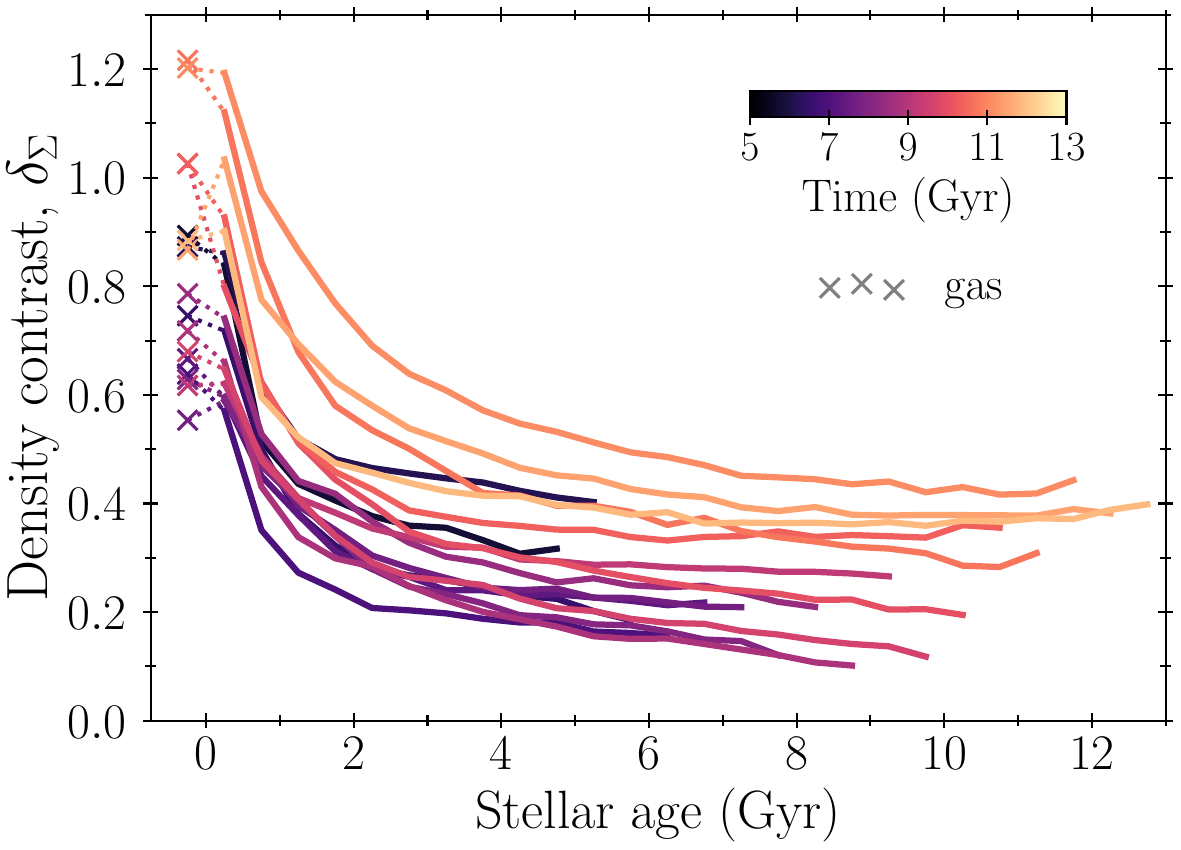}
        \caption{Spiral arms density contrast for different populations and times. The lines show the relative ratio between the mean mass surface density of the arms and the disc average mass surface density as a function of stellar age. Each line corresponds to a different time in the simulation (with darker meaning earlier, as in Fig.~\ref{fig_surfdenscontrasts}). Crosses linked with dotted lines to guide the eye correspond to the gas density contrast at the corresponding time. We exclude galactocentric radii smaller than 4\,kpc and larger than 8\,kpc. }
        \label{fig_agehist}
    \end{figure}
    \begin{figure}
        \includegraphics[width=0.48\textwidth]{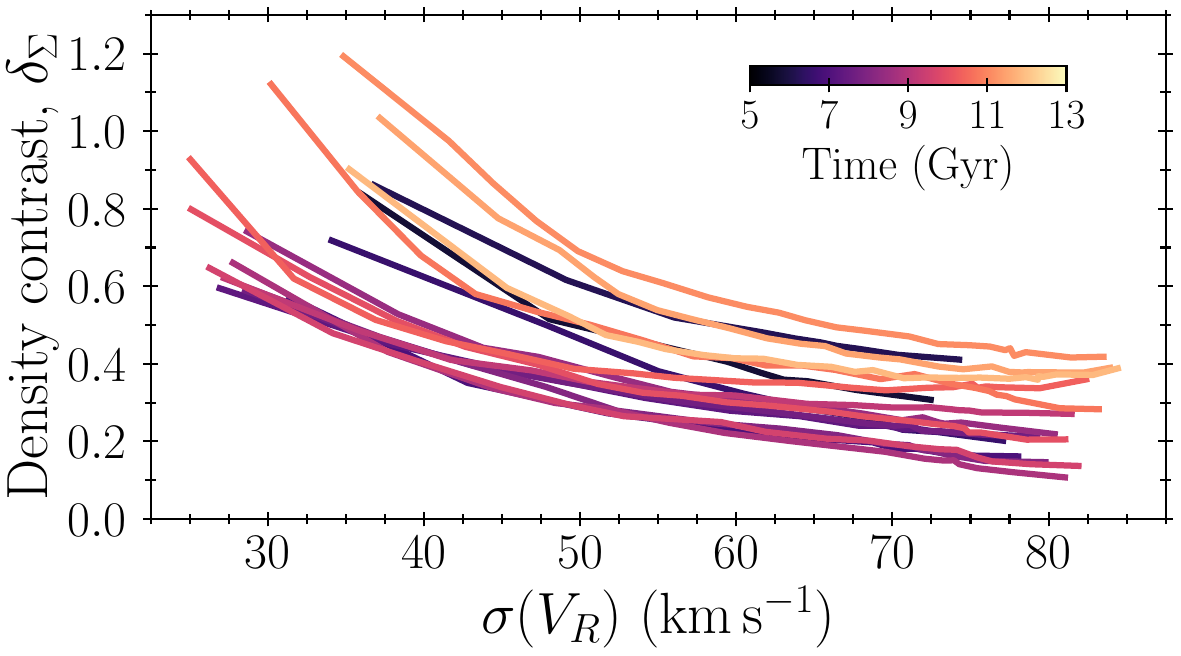}
        \caption{Spiral arms density contrast for different radial velocity dispersions and times. Each line corresponds to a different time in the simulation (with darker meaning earlier, as in Fig.~\ref{fig_surfdenscontrasts}). We consider only galactocentric radii between 4 and 8\,kpc.}
        \label{fig_veldisp}
    \end{figure}
    
    In Fig.~\ref{fig_agehist} we show how the density of the spiral arms compared to the average disc depends on the age of the stellar populations and on time. We also show the density contrasts for gas, marked as crosses. Lines of different colours show measurements from different snapshots of the simulation, with each snapshot including progressively older stellar populations as time advances.
    The figure shows that the spiral arms in the youngest stars and gas have similar density contrasts and present the largest contrasts  of all populations at all times. Newly formed stars can present contrasts of up to 120\% as a result of a gas distribution very concentrated in arms.
    At any given time, $\delta_{\Sigma}$ then decreases significantly and rapidly with stellar age until it reaches a plateau after a few billion years.
    The global amplitude of these curves depends on the global strength of the arms. We see a behaviour very similar to the global spiral structure in Fig.~\ref{fig_surfdenscontrasts}, especially when comparing the values of the plateau of the oldest stars: the spiral structure is globally stronger at the beginning and end of the timespan analysed.
    Besides this, $\delta_{\Sigma}$ is always positive for all ages, indicating that spiral arms are ubiquitous in all populations. Interestingly, stars as old as $\sim$\,12\,Gyr are organized in arms that have contrasts around 40\% at the latest times of the simulation, when the arms are strong.

    The different age populations react and contribute to the spiral structure based on their kinematic temperature. To further check this connection, in Fig.~\ref{fig_veldisp} we show the dependence of the density contrasts on the radial velocity dispersion of stars, $\sigma(V_R)$. For this plot, we calculated the standard deviation of $V_R$ for different stellar age intervals within the 4-8\,kpc ring as representative values for the velocity dispersion\footnote{As expected, $\sigma(V_R)$ decreases with galactocentric radius and increases with stellar age \citep[Fig.~10 in][]{Debattista24_sim}.}. In this figure we see that, as expected, at a given time smaller velocity dispersions correspond to larger spiral arm density contrasts. However, a given velocity dispersion can correspond to a large range of density contrasts. This scaling of curves is similar to the one in Fig.~\ref{fig_agehist}, which is connected to the overall spiral arms strength at a given time \footnote{In fact we also observe how at the times when spiral arms are stronger, the radial velocity dispersion of a given population grows, possibly due to the strong streaming motion and migration associated with the spiral arms.}. The similarity between Figs.~\ref{fig_agehist} and \ref{fig_veldisp} shows a strong connection between stellar ages and velocity dispersions, and their corresponding spiral arm strengths.
    
\begin{figure}
        \includegraphics[width=0.5\textwidth]{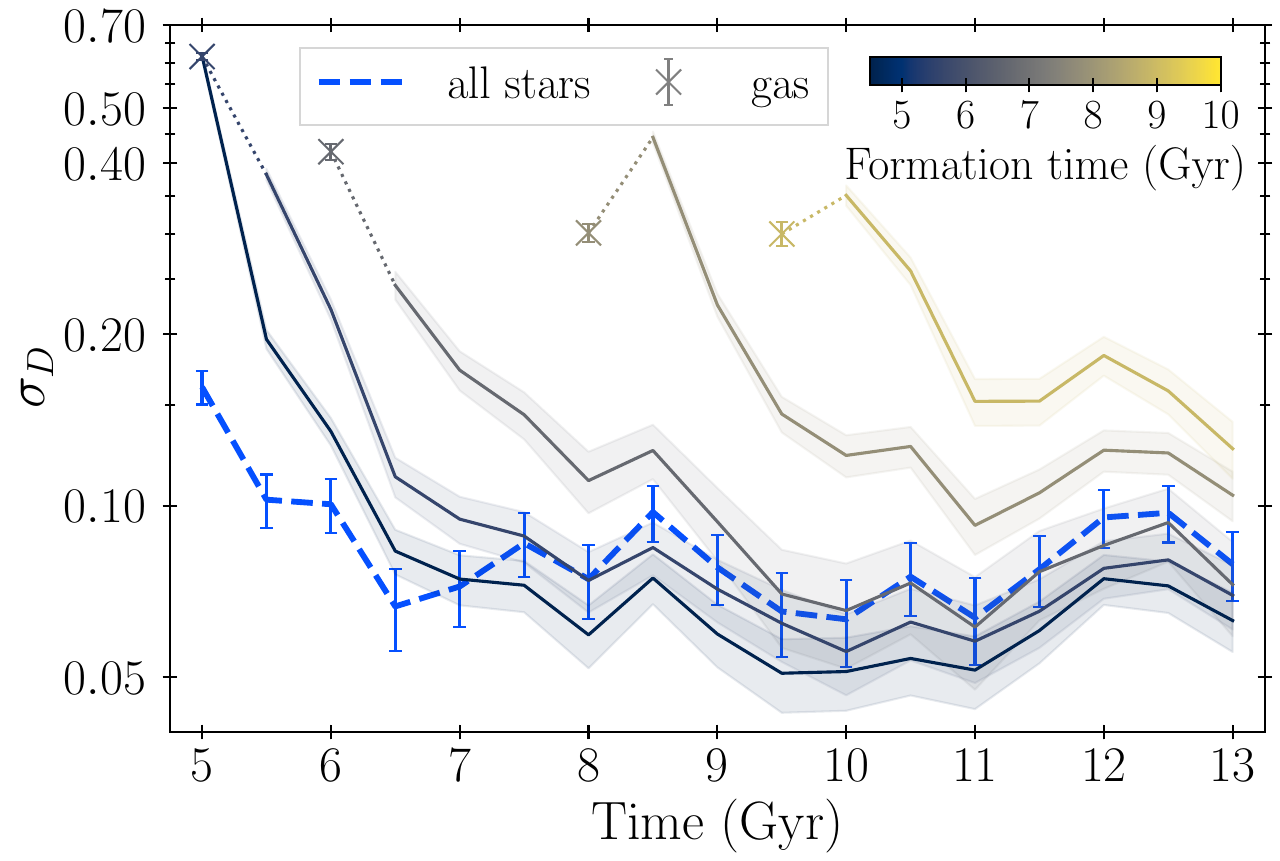}
        \caption{Temporal evolution of the standard deviation of $D$, $\sigma_D$, for several co-forming groups of stars traced along snapshots (solid lines) and for all stars together (dashed line). Crosses correspond to the gas of the snapshot 0.5\,Gyr before the formation of each group and are linked to the curves with dotted lines to guide the eye. Shaded areas show the dispersion for each formation time due to short-term fluctuations, computed as the mean standard deviation of $\sigma_D$ from intervals with higher-cadence. The innermost 4\,kpc and radii beyond 8\,kpc are excluded.}
        \label{fig_Dtform}
    \end{figure}

    We now examine the level of spiral structure of the stellar disc for selected groups of stars born at a given time  and how it varies with time. In contrast to examining stars of a particular age at a given time as before, this helps us analysing the temporal evolution of the response to the spiral arms of fixed populations. For this, we use our other quantifier of the level of structure: the standard deviation of the $D$ values, $\sigma_D$.
    In Fig.~\ref{fig_Dtform} we show the temporal evolution of $\sigma_D$ of all stars (dashed line) and of particular sets of stars born at certain times (thinner solid lines). These sets are comprised of stars born during 0.5\,Gyr-width intervals starting at 4.5, 5, 6, 8 and 9.5\,Gyr, and we track them until the end of the simulation. The dispersions of $D$ computed from the gas previous to the formation time of the different sets (0.5\,Gyr before) are also indicated by crosses. As before, we calculate $\sigma_D$ only for the region within 4 and 8\,kpc to exclude the bar and the outskirts of the disc.
    We see that, as expected, the gas shows $\sigma_D$ values that are similar to the stars born immediately after. Globally, as time progresses and the traced stars become older, the dispersion of $D$ values gradually decreases (as also seen in Figs.~\ref{fig_Dkde} and \ref{fig_DandDensity}). This indicates again that the spiral arm strength decreases as a population becomes older.
    Besides this global trend, at certain times $\sigma_D$ increases consistently in all selections of stars born along the entire evolution and also for stars of all ages together (especially evident after 11\,Gyr). While the 0.5\,Gyr temporal baseline used for the analysis and the transient nature of the spiral arms may be responsible for the small-scale sharp gradients in $\sigma_D$, the consistency among the different selections is indicative of a true increase of the global strength of the spiral arms as in the analysis above. 

    As discussed later (Sect.~\ref{sectDiscussion}), the strengthening of arms after 11\,Gyr could be related with the bar weakening around that time.
    That said, irrespective of the origin of this strengthening, from this section we conclude that old populations present weaker spiral structure in general but they do not lose their ability to respond constructively with a strong spiral perturbation. Indeed, old stars can reinforce their spiral structure and present more structure than they previously did if the overall spiral strengthens, which implies that there is no single scaling between spiral strength and velocity dispersion.

% ------------------------------------------
\subsection{Temporal association of stellar populations with arms}\label{sectTrace}

    With the aim of understanding in more detail the global link between spiral arm strength and velocity dispersions or ages, in this section we investigate the dynamics of various stellar populations, focusing on their role in shaping and responding to spiral arms.
    To this end, we trace the positions of stars of different ages with respect to the arms outlined by the local dimension contours with the high temporal cadence (5\,Myr) available for our model.
    
    \begin{figure*}[h!]
    \begin{center}
        \includegraphics[width=0.95\textwidth]{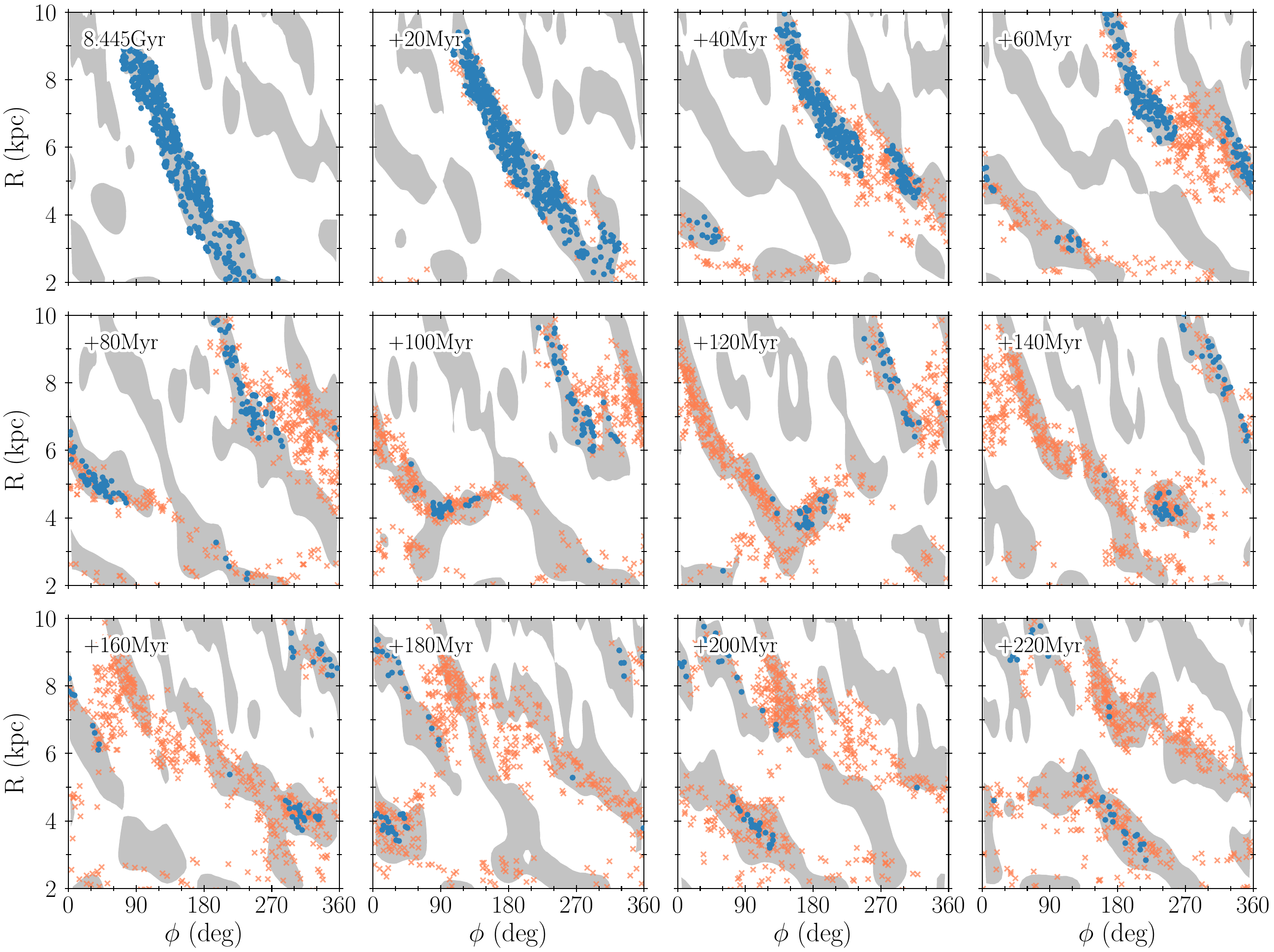}
        \caption{Evolution of particles born within the spiral arm around $\phi$\,$\sim$\,$180^{\circ}$ at time 8.445\,Gyr. Grey shaded areas show spiral arms of stars younger than 0.5\,Gyr defined by $D$\,$\leq$\,1.9. Blue dots are stars that have remained inside arm regions, while orange crosses show stars having left the arms region at least once. The disc rotates in the direction of increasing azimuth.}
        \label{fig_diffusionFollow}
    \end{center}
    \end{figure*}

    Figure~\ref{fig_diffusionFollow} presents a sequence of snapshots in which stars born inside a spiral arm at 8.445\,Gyr are monitored over a 220\,Myr period.
    Stars are initially represented with blue dots and are replaced by orange crosses when they leave the arm for the first time.
    The outline of the spiral arms (as defined in Sect.~\ref{sectMethod}) is shown as grey-shaded regions.
    Visual inspection of Fig.~\ref{fig_diffusionFollow} shows that the young stars born in the spiral arm remain in it until the arm breaks into smaller segments, which will in turn reconnect to adjacent structures shortly later. In this example, this reconfiguration of the arms occurs in less than 100\,Myr, which is a typical timescale that we find also at other times and for other arms, and which agrees with the lifetimes of spiral arms formed in other simulations \citep{Wada11_noOffsets, Roskar12_shortlivedArms,Pettitt20_ISMdiffModels}.
    During this time of about 100\,Myr (seen in the first 4--5 panels of Fig.~\ref{fig_diffusionFollow}), the initial arm gets stretched, breaks and then reconnects to another arm segment.
    The traced stars (the large majority of which have now become orange crosses) remain on this new arm, until it reorganizes again.
    This continuous rearrangement of the spiral structure may be explained by the interaction of multiple spiral patterns that repeatedly detach and reattach generating ever-changing spiral arms \citep[as in the case of a bar interacting with arms, e.g.][]{Sellwood88_multipleModes}.
    Independent of this, this analysis shows that young stars, which are born preferably on top of spiral arms, quickly become associated to other arms after the spiral arms break. Interestingly, about 7\% of stars remain coincident with the spiral structure even after 220\,Myr (blue dots in the last panel), meaning they never leave spiral arms during this period.

    \begin{figure}
    \begin{center}
        \includegraphics[width=0.45\textwidth]{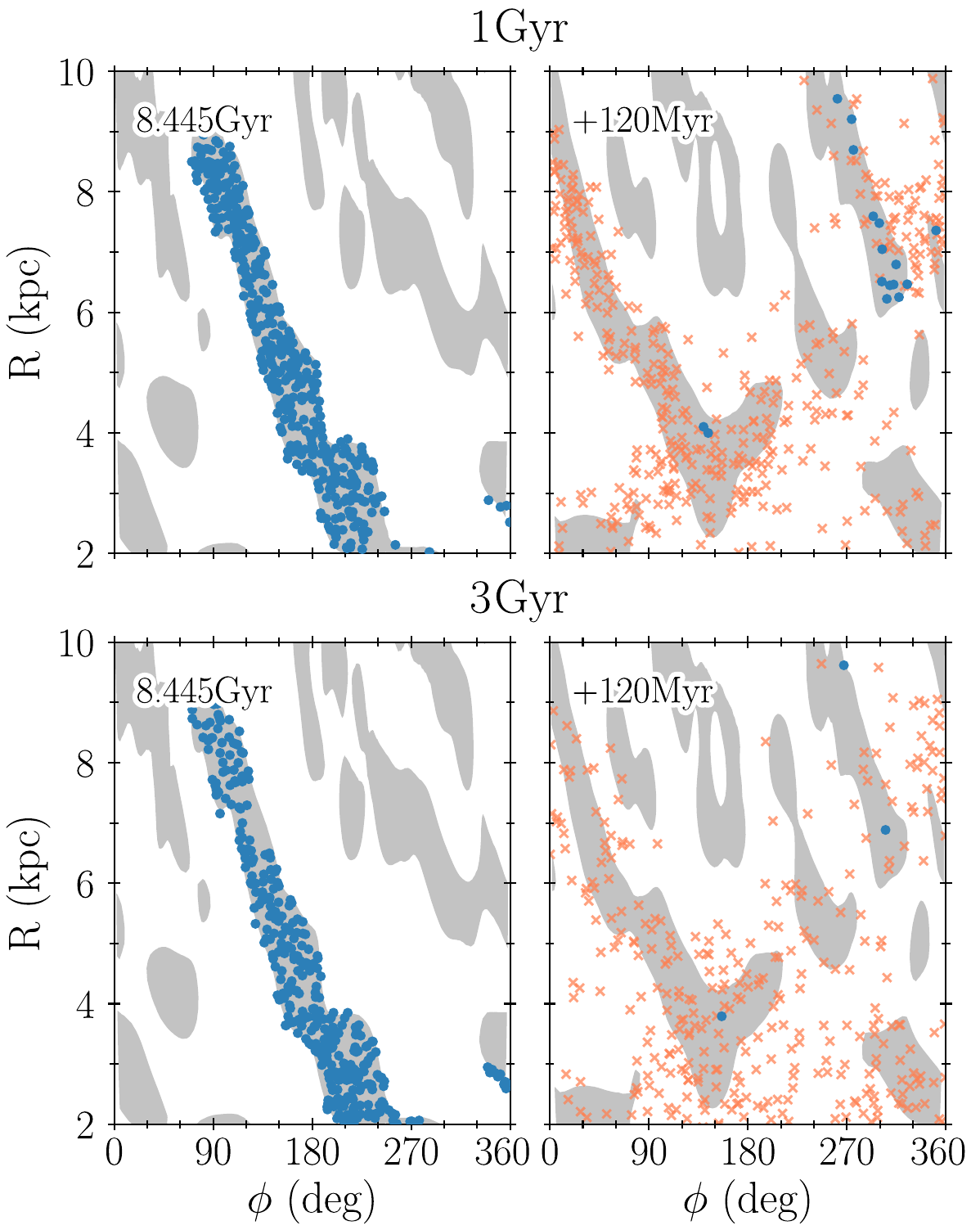}
        \caption{Same as Fig.~\ref{fig_diffusionFollow}, but showing just two snapshots with stars having initial ages 1--1.005\,Gyr (top) and 3--3.005\,Gyr (bottom). The starting arm in the left panels evolves into the upper right arm segment containing most of the blue dots in the right panels after a half rotation.}
        \label{fig_diffusionFollowOld}
    \end{center}
    \end{figure}

    While Fig.~\ref{fig_diffusionFollow} traces the position of the youngest stars born in a given arm at the given time, we now focus on older stars (ages 1--1.005\,Gyr and 3--3.005\,Gyr) that were coincident with the same arm as in the previous example. We look at how their positions change with time in Fig.~\ref{fig_diffusionFollowOld}.
    Most $\sim$\,1\,Gyr-old stars (top panels) are still found in the spiral structure after 120\,Myr (but not in the same arm, since the initial arm is now the upper right segment where some blue points still remain). On the other hand, whereas the $\sim$\,3\,Gyr-old stars (bottom panels) still overlap with the spiral arms contours after the same time period, they have a wider distribution.
    
    The examples of Figs.~\ref{fig_diffusionFollow} and \ref{fig_diffusionFollowOld} show that the older the stars get, the harder it is for them to find their way to a new arm after the initial arm in which they were located changes. In Appendix~\ref{app_traceArms} we show another example of this same  transition between stellar ages at another time in the simulation.

    \begin{figure}
    \begin{center}
        \includegraphics[width=0.48\textwidth]{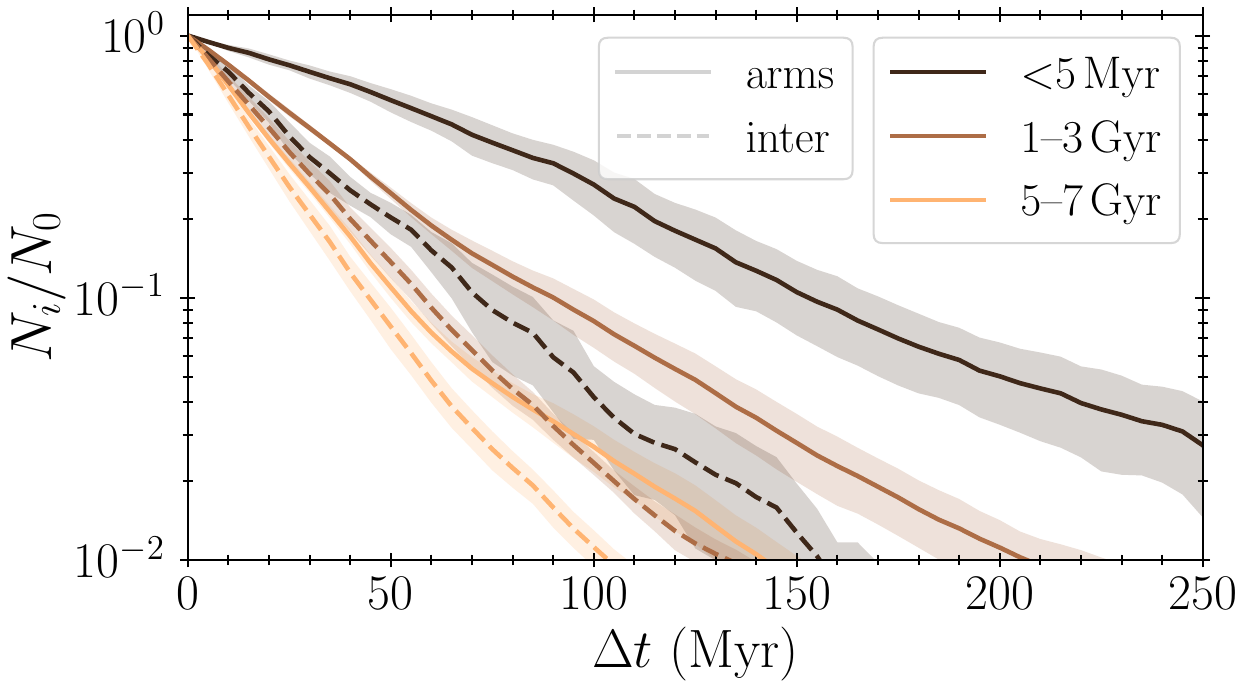}
        \caption{Fraction of stars within spiral arms ($D$\,$\leq$\,1.97, solid) or interarms ($D$\,$\geq$\,2.06, dashed) as a function of time elapsed since their initial selection. We include times between 8.25 and 8.75\,Gyr. Dark lines trace particles that are born at the initial time, while lighter ones trace particles that are 1--3\,Gyr-old and 5--7\,Gyr-old initially.}
        \label{fig_diffusionLeave}
    \end{center}
    \end{figure}

    To further quantify this association of stars of different ages with the spiral structure, we now study the fraction of stars located in spiral arms as a function of time. We start by selecting the initial set of $N_0$ stars born (or located) in any arm within the radial range of 6--8\,kpc at a given time (this is narrower than our earlier radial range to limit the smearing out due to shear). We trace them over time and calculate the fraction $N_i/N_0$, where $N_i$ is the number of stars that are still coincident with spiral arms at each time (this would correspond to points that remain blue in Figs.~\ref{fig_diffusionFollow} and \ref{fig_diffusionFollowOld}, but for the whole azimuthal range instead of just one arm). We repeated the same procedure starting from six equally spaced times between 8.25 and 8.75\,Gyr, and averaged the resulting fractions to avoid any special behaviour from a single time.
    We did this exercise for very young stars born on the arm (younger than 5\,Myr at the beginning), and for intermediate (1--3\,Gyr-old) and old (5--7\,Gyr-old) stars.
    We also perform an analogous analysis  but for stars selected in the interarm regions instead of the arms. To obtain a fair comparison, in this case we adapted the $D$ threshold so that the area covered by interarms is as close as possible to the area covered by arms: we used $D^-_{\mathrm{th}}$\,=\,1.97 to define arms (where $D$\,$\leq$\,$D^-_{\mathrm{th}}$) and $D^+_{\mathrm{th}}$\,=\,2.06 to define interarms (where $D$\,$\geq$\,$D^+_{\mathrm{th}}$). Values in-between are considered boundaries not belonging to arms or interarms.
    
    This mean fraction of stars remaining within the arms as a function of the time elapsed since the initial selection is shown as solid lines in Fig.~\ref{fig_diffusionLeave} for the three ages chosen. We show the results for the interarm regions with dashed lines.
    Uncertainties are computed by adding in quadrature the noise from the number of particles at each step modelled as an exponential decay and the dispersion due to different starting snapshots. The latter usually dominates, while the former becomes relevant for samples with very few stars, such as when considering the fractions of very young stars abandoning interarm regions.
    
    \begin{table}[]
        \centering
        \caption{Timescales (in Myr) to reduce from 100\% to 10\% the number of stars initially selected within the arms and in interarms (labelled `inter') at times between 8.25 and 8.75\,Gyr.
        Different rows give the values for different galactocentric rings and each pair of columns indicates a population with a different age.
        Uncertainties are of the order of 5--20\,Myr, mainly depending on age and whether the initial selection is in arms or in interarms (Fig.~\ref{fig_diffusionLeave}).}
        \begin{tabular}{c|cc|cc|cc}
                   & \multicolumn{2}{c|}{<\,5\,Myr} & \multicolumn{2}{c|}{1--3\,Gyr} & \multicolumn{2}{c}{5--7\,Gyr} \\
                   &      arm      &     inter     &      arm      &     inter     &      arm      &     inter     \\ \hline
        4--6\,kpc  &      141      &       48      &       65      &       51      &       42      &       41      \\
        6--8\,kpc  &      158      &       73      &       90      &       58      &       53      &       45      \\
        8--10\,kpc &      184      &       68      &       96      &       61      &       56      &       47     
        \end{tabular}
        \label{timescales1650}
    \end{table}

    Table~\ref{timescales1650} gives the timescales for different populations to leave spiral arms or interarm regions obtained from fitting an exponential decay to Fig.~\ref{fig_diffusionLeave} (forcing the initial fraction to be 100\% and restricting to fractions larger than 5\%). We define these timescales as the time needed to lose 90\% of the initial selection\footnote{Older stars may have already been within the arm or interarm region before our initial selections, which means that their timescales may be slightly underestimated. However, since observed trends also hold between intermediate age and old stars, we expect this effect to be small enough for our conclusions to remain unchanged.}.
    The table also includes two more radial rings, showing that our conclusions do not depend strongly on galactocentric radius.
    It is important to keep in mind that these timescales should only be related with the departure from the initially selected spiral arms or interarm. Stars still participate in the global spiral structure for longer time intervals, although more loosely as they age.

    Different times in the simulation yield slightly different timescales depending on the characteristics of the spiral structure at those times, but the general trends of Fig.~\ref{fig_diffusionLeave} and Table~\ref{timescales1650} are maintained.
    First, as expected from the increase of velocity dispersion with stellar age, the older the stars, the shorter their association with the arms. Typical timescales in which half of the just born stars leave the arms are 50--70\,Myr and almost all of them require less than 140--180\,Myr to do so. This is similar to the interval between arm reconnections and increases slightly with radius (perhaps because of the radial dependence of the dynamical timescale).
    Old stars leave arms about 3 times faster than young ones.
    Second, young stars also leave interarm regions more slowly than old ones, but differences are not so large in this case.
    Last, although the arms are complex and transient, all populations tend to spend more time within spiral arms than in interarms, showing the effects of the gravitational perturbation of the arms. This is true especially for very young stars, where these two timescales can differ by a factor of 2--3.
    
    Many stars leave arms when they break and then rejoin another arm segment, while some stars remain continuously within arm-like regions for more than 200\,Myr (3--7\% of the youngest stars). It is not straightforward to unambiguously explain these two simultaneous behaviours within the framework of a particular theory of spiral structure.
    Despite that, we show here that the association between stars and spiral arms in our model depends on the population. Kinematically cold young stars are more likely to participate for longer in arm overdensities, even when their pattern is evolving rapidly in a complex way.

% ------------------------------------------
\subsection{Torques exerted by different stellar populations}\label{sectTorques}

    One of our previous results is that young stars present the strongest spiral structure but the oldest stars show non-negligible arms as well. We now investigate which population contributes the most to the non-axisymmetric forces due to spiral structure: young stars --which have more overdense and stronger spiral arms-- or old stars --which are more numerous?
    To answer this, we analysed the torques exerted on the 4--8\,kpc ring by different stellar populations. We first calculated the potential created by a given stellar population and derived the corresponding force per unit of mass using \textsc{Agama} \citep{2019MNRAS.482.1525V}. We then computed the in-plane torques in a grid inside the aforementioned ring. Since we want to avoid the torques created by the bar, we azimuthally shuffled the particles inside 4\,kpc before obtaining the potential. Interestingly, we find that our results were very similar when we used the original azimuths (without shuffling), indicating that between 4 and 8\,kpc the non-axisymmetric forces due to the spiral arms dominate over those created by the bar.  Figure~\ref{fig_torquesXY} in Appendix~\ref{app_torquesmaps} shows examples of torques maps.
    
\begin{figure}
        \includegraphics[width=0.5\textwidth]{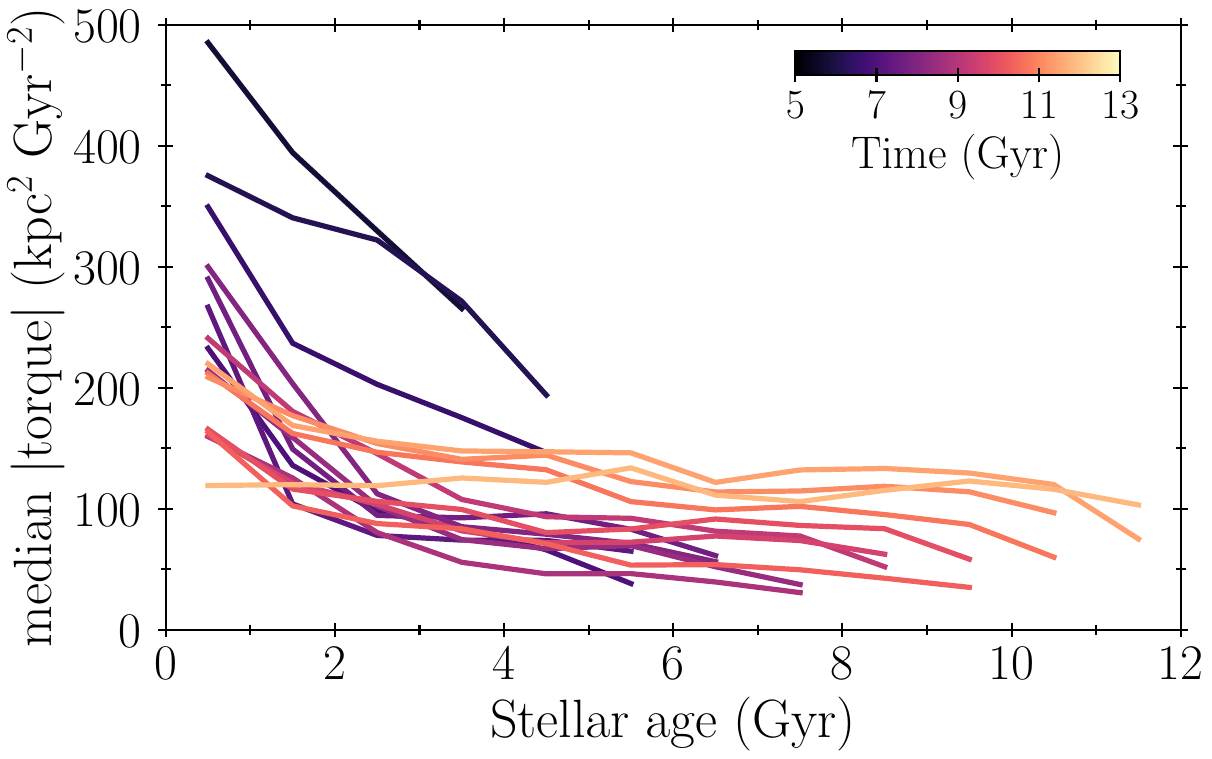}
        \caption{Torques due to spiral arms for different populations and times. The lines show the median absolute values of torques (exerted within 4 and 8\,kpc by stars at any disc location) as a function of stellar age. Each line corresponds to a different time in the simulation as in Figs.~\ref{fig_agehist} and \ref{fig_veldisp}.}
        \label{fig_torques}
    \end{figure}

    Figure~\ref{fig_torques} shows the age dependence of the median absolute value of those torques at different times of the simulation.
    At early times (darkest curves), there are very strong torques showing a steep decrease with stellar age. This is due to a large number of young stars with large spiral arm contrast and relatively large pitch angles (seen in Fig.~\ref{fig_DandDensitytime}). At intermediate times, when the arms are relatively weak, we see globally low torques that decrease more slowly with stellar age. Finally, at the lattermost times (lightest curves), we see strong torques that are quite constant with stellar age. Between times 12.5 and 13\,Gyr, stars older than 5\,Gyr contribute with similar torques as stars younger than that.
    This might be because old stars largely exceed young ones in numbers at these times, even if old ones have smaller density contrasts, and also because the properties of the spiral structure itself might have changed.
    Indeed, the torques produced by a specific population decrease as it ages at early times and then increase at late times, indicating that arm strength evolves in the same way, as we already discussed in the previous sections.
    
    In conclusion, when considered individually, young populations generate stronger torques than old populations on average but the bulk of intermediate age and old stars are the ones contributing the most overall to the torques because they are more numerous. In addition, the particular contribution of each population changes with time due to the different factors mentioned above.
    At the later times of the simulation, we find almost equal contribution to the torques from all stellar ages and we see that the spiral structure made of old stars produces even larger torques than younger ones at other times with weaker arms.

%--------------------------------------------------------------------
\section{Discussion}\label{sectDiscussion}

    One of our main results is the presence of spiral arms in absolutely all populations and their different reaction to the arms depending on their age.
    Gas and very young stars present stronger arms than old stars (Figs.~\ref{fig_DandDensity}, \ref{fig_agehist} and \ref{fig_Dtform}), which is consistent with other simulation studies \citep[e.g.][]{Khoperskov18_ArmsFeHkinPops,Ghosh22_oldSimArms,Debattista24_sim} or observational data, both in the Milky Way \citep[e.g.][]{Poggio22_ArmsChemistry, Drimmel22_assymDiscDR3} and in external galaxies \citep[e.g.][]{Chandar17_M51armsOC, Shabani18_offsetsObs}.
    \cite{Ghosh22_oldSimArms} analysed the same simulation that we use here at time 12\,Gyr and also found spiral arms for all stars (even in their old age-interval between 8 and 12\,Gyr). Here, thanks to the resolution of this model combined with our new method, we extend their result detecting spiral structure in all populations throughout the evolution of the model and using narrower stellar age intervals. At late times in the simulation we see spiral arms even in stars as old as 11--12\,Gyr and with radial velocity dispersions larger than 70\,km\,s$^{-1}$.
    
    As discussed in the introduction, some studies have not found spiral structure in old stellar populations in their models. This can be caused by either a different, rough detection-method or different physical state of the model, such as very high velocity dispersions, which would hinder spiral arm formation. To better know whether it is the former or the latter, we analysed, using our methodology, the publicly available g2.79e12 NIHAO-UHD model \citep{Buck20_NIHAO} and found, similarly to \cite{Palicio25_oldSimArms}, that spiral structure is absent in stars older than about 3\,Gyr. This model has more flocculent structure than ours and a much hotter stellar disc (radial velocity dispersions above 100\,km\,s$^{-1}$ for stars older than 3\,Gyr). This kinematic temperature is likely one of the main reasons for developing weak arms, which also might not be strong enough to engage the hotter populations.
    On the other hand, for models with very strong spiral arms, \cite{Bernet25_DarkSpirals} recently found that they can leave clear spiral imprints in the dark matter halo, implying that even dispersion supported populations can feel the influence of the spiral arms. In the case of the halo, though, the effect occurs through a gravitational wake, which is likely not the same process that acts on the old (but relatively cold) stellar disc in our simulation.
    
    Throughout the evolution of our simulated disc, velocity dispersions are in the range 15--60\,km\,s$^{-1}$ and 25--85\,km\,s$^{-1}$ in the vertical and radial directions, respectively, depending on radius and stellar age. These ranges are similar to the values obtained for the Milky Way \citep{Sanders18_AVR,Anders23_AVR, McCluskey25_AVR}.
    While the detection of spiral arms in intermediate age stars in the Galaxy has been reported  \citep[e.g.][]{Lin22_MWarmsRC,Uppal23_MWarmsRC,Khanna24_oldMWarms}, our results emphasize that even older stars could be organized in spiral structure if the arms are strong.

    The role of all stellar populations in shaping and responding to spiral arms is further supported by our finding that stars of all ages spend more time within arms than in interarms (Sect.~\ref{sectTrace}). The dependency of the spiral arms on the kinematic temperature and hence on the age is also seen in the fact that old stars effectively participate in spiral arms for shorter times than young stars (i.e. kinematically hotter populations, with larger epicycles, leave arms faster than colder ones). These aspects have been possible to quantify thanks to the high cadence (5\,Myr time-steps) of the simulation, which has little precedence, and the ability to precisely outline the arm locations with our new method. Our results can also help to explain the findings that spiral arm regions present lower average age and metallicity \citep{Khoperskov18_ArmsFeHkinPops,SanchezMenguiano20_ExtArmsChem,Poggio22_ArmsChemistry,Hawkins23_ArmsChemistry,Chen24_MAGPI2gal,Breda24_externalArmsAVR,Hackshaw24_ArmsChemistry,Barbillon25_ArmsChemistry,Debattista24_sim,ViscasillasVazquez25_ArmsChemistry}: apart from newly formed stars being preferentially born in spiral arms, they spend more time in them and quickly get rearranged into the evolving new spiral structure.
    In addition, as we see that populations with different kinematic temperatures may move with respect to arms in different ways, we speculate that this can have implications on the amount of radial migration for different populations, which in turn is related to the distribution of chemical elements at galactic scales. 

    Despite the weakening of the spiral structure as the age and velocity dispersion of a population increase, we also find that the density contrast and the level of substructure can increase at specific times for all populations, including for old and kinematically hot stars. This reinvigoration of spiral arms in old stars coincides with the strengthening of arms globally in the disc.
    Similarly, we find that the contribution of different stellar populations to the torques exerted on the disc also changes with time (Fig.~\ref{fig_torques}), which has been calculated here for the first time, to our knowledge. This is related to the mass contribution of each population at each time and to the evolution of the global spiral structure itself, up to a point where old stars can produce torques as strong as those from young ones at the late times of the simulation.

    The revitalization of spiral arms in this isolated model at later times is an interesting phenomenon that can be related to several aspects of the disc evolution. As the star formation is continuously ongoing in this simulation, the disc mass is building up with time, which can affect the spiral instabilities. Moreover, the bar weakens with time.
    A sudden bar damping may create strong spiral structure, as described in \citet{Lokas16_BarDamping}. The destruction of the bar can also change the gas flows in the disc, and the arms can be amplified by the gas that previously was funnelled by the bar. Furthermore, the lowering of vertical velocity dispersions towards the end signals that the disc is in a different kinematical state, which can affect the spiral structure as well. As the disc grows inside-out, its density profile changes, which can also include the position of breaks. If the spiral arms in this model are related to edge modes \citep{Fiteni2024}, the change of the position of the density break could affect the strength of the spirals in the fixed ring in which we analyse them. While it is not our goal to determine the exact reason for this global evolutionary sequence of the spiral arms, we conclude that this evolution is accompanied by changes in the engagement of different populations with the arms, and in the relative torques they create.

    Finally, we highlight some future directions. Our methodology can be easily applied to other simulations to check whether our results hold in different prescriptions for star formation and in different arm models (e.g. tidally induced arms).
    The local dimension can also be applied to Milky Way stellar samples and to other local galaxies with resolved stellar populations.
    Furthermore, it could also be adapted to continuous photometric distributions from external galaxies. For example, it could be interesting to study the connection between the strength of the spirals arms and their torques from different populations, and the evolutionary stages or properties of galaxies in observations. Indeed, we could in theory compute torques due to different stellar populations in external galaxies from multi-wavelength photometry and assuming a mass-to-light ratio \citep{Gnedin95_torquesArms}.
    Thus, our new tool would allow for both observations and simulations to be analysed in a comparable way. Another natural step could be studying the detailed arms dynamics thanks to the robust delimitation of the arms location with our method, including a systematic analysis of stellar orbits and other spiral arms parameters (such as pitch angle, or chemical and kinematic patterns).

%--------------------------------------------------------------------
\section{Summary and conclusions}\label{sectConclusions}
    
    In this paper we present a new tool\footnote{We will make the code available upon acceptance.}
    to outline spiral arms from the geometry of the density distribution of a galaxy. Our method measures the local dimension of the distribution around a given point.
    This method is independent of discrete bases, models or user guidance (does not rely on Fourier expansions, kernel functions, logarithmic fittings or by-eye tracing), and it allows us to delimit the spiral arms with high precision and to highlight low-amplitude overdensities. We have applied this method to study the spiral structure across populations in the isolated galaxy of the M1\_c\_b model from \cite{Fiteni21_sim}. This is a high-resolution hydrodynamical simulation that ends with a disc of more than 11 million stellar particles formed from gas cooling down from a corona onto a disc.
    The results of our analysis can be outlined as follows:

    \begin{itemize}
        \item Stars of all ages (even 11--12\,Gyr-old stars) participate in creating spiral structure.
        \item The spiral pattern is similar across ages, but with young stars exhibiting more substructure.
        \item Spiral arm strength decreases with age and with velocity dispersion, which are correlated. This decrease occurs especially for stars younger than 2\,Gyr, while the density contrasts are quite similar among older stars.
        \item The existence of spiral arms in the old and kinematically hot populations and the trend with age are further evidenced by the finding that stars of all populations spend more time in spiral arms than in interarm regions but young stars remain within spiral structure for longer.
        \item Almost all young stars remain coincident with a given arm at most 140--180\,Myr, depending on radius. This time decreases to 40--60\,Myr for old stars. Newly formed stars remain aligned with the arm where they form until it breaks or reconnects with other arms and they soon rejoin in another segment of the spiral structure; the overlap with arms persists as the stellar age increases, but more loosely.
        \item Despite the general trend with age, there is a large time variability in arm strength: a given velocity dispersion (or age) can correspond to a wide range of spiral strengths depending on the global spiral structure at that time. This, in turn, could depend on the bar strength and on how the disc evolves and its mass builds up.
        \item This variability translates into stars younger than 2\,Gyr having density contrasts between 20\% and 120\%, depending on time; while contrasts for old or kinematically hot populations are between 10\% and 50\%.
        \item Despite spiral arms being stronger in young populations, the bulk of the spiral non-axisymmetric torque, which dominates over bar torques at intermediate radii in this model, originates from populations of different ages, depending on the simulation time. At early and intermediate stages of the simulation, young populations (<\,2\,Gyr) dominate the non-axisymmetric forces individually, whereas at later times, stars of all ages contribute equally to the torques.
        \item The velocity dispersions of the oldest populations in the model are comparable to those of old Milky Way disc populations, suggesting that, if the arms of our Galaxy are similar to those in this model, spiral structure could also be present in old Galactic populations.
    \end{itemize}

    Our results demonstrate the high potential of the local dimension method to detect spiral arms.
    We conclude that spiral arms are not exclusive to young tracers, but rather are global features that affect all the components of a galactic disc.
    Fully understanding the origin and nature of spiral arms thus requires accounting for the contribution of all stellar populations, an effort that becomes increasingly crucial in the era of large stellar samples for the Milky Way and of integral field unit data for external galaxies.

%%%%%%%%%%%%%%%%%%%%%%%%%%%%%%%%%%%%%%%%

\begin{acknowledgements}
    V.P.D. is grateful for the hospitality during his visit to the University of Barcelona. \\
    This research made use of
    \texttt{python} \citep{python}, Astropy\footnote{\url{http://www.astropy.org}} \citep{astropy:2013, astropy:2018},
    \texttt{matplotlib} \citep{Hunter:2007},
    \texttt{Numba} \citep{numba:2015, Numba_5847553},
    \texttt{numpy} \citep{numpy}, 
    \texttt{pynbody} \citep{Pynbody:2013,Pynbody13:2021} and \texttt{scipy} \citep{2020SciPy-NMeth, scipy_10155614}.
    Software citation information aggregated using \texttt{\href{https://www.tomwagg.com/software-citation-station/}{The Software Citation Station}} \citep{software-citation-station-paper, software-citation-station-zenodo}.\\
    This work was part of the PRE2021-100596 grant funded by Spanish MCIN/AEI/10.13039/501100011033 and by ESF+.
    It was also (partially) funded by "ERDF A way of making Europe" by the “European Union” through grants RTI2018-095076-B-C21 and PID2021-122842OB-C21, and the Institute of Cosmos Sciences University of Barcelona (ICCUB, Unidad de Excelencia ’Mar\'{\i}a de Maeztu’) through grant CEX2019-000918-M.
\end{acknowledgements}

%%%%%%%%%%%%%%%%%%%%%%%%%%%%%%%%%%%%%%%%

\bibliography{references}

%%%%%%%%%%%%%%%%%%%%%%%%%%%%%%%%%%%%%%%%

\begin{appendix}
\section{Methodological details of the local dimension analysis}\label{app_Dfits}

\begin{figure}[h]
        \includegraphics[width=0.47\textwidth]{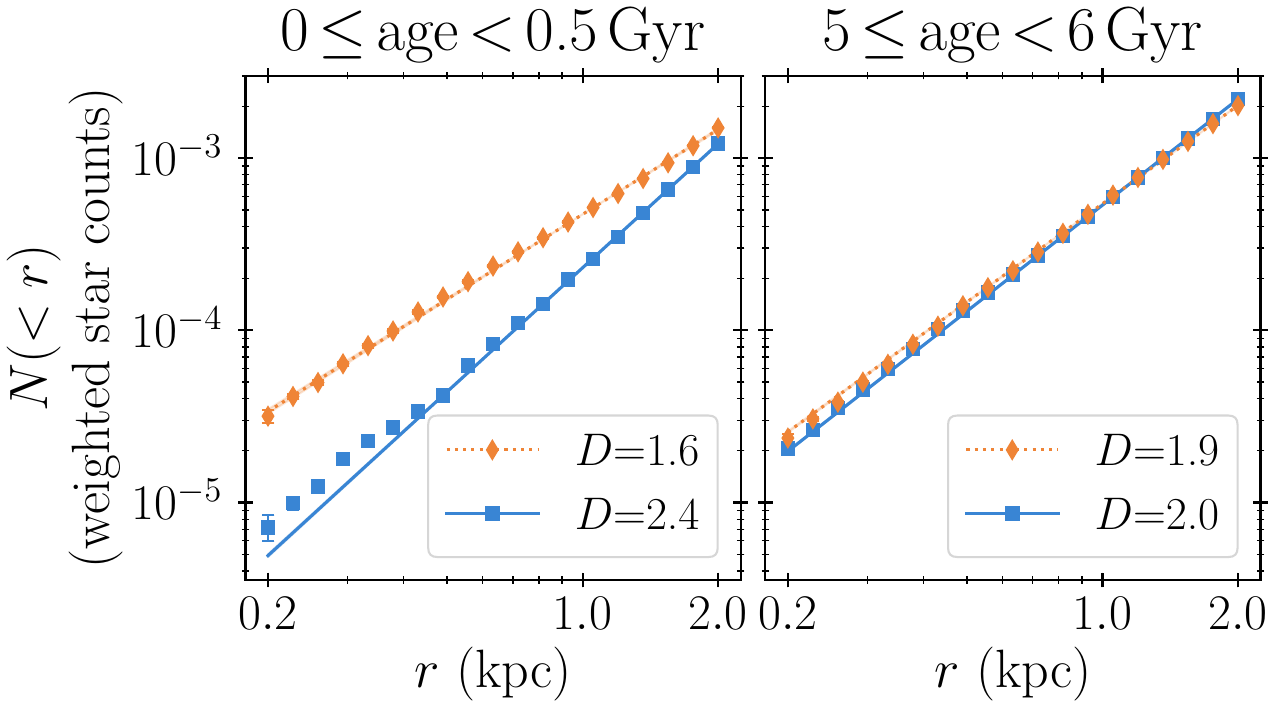}
        \caption{Regressions to obtain the local dimension, $D$, at two galactic locations from the number of particles, $N$, closer than a distance $r$ to the given location. The left panel considers stars younger than 0.5\,Gyr while the right one uses 5--6\,Gyr-old stars. Each panel shows the cases of an arm (orange dashed line and diamonds) and an interarm region (blue solid line and squares).}
        \label{fig_DfitsApp}
    \end{figure}

    In this appendix we describe some methodological aspects of the local dimension and parameter choices. First, in Fig.~\ref{fig_DfitsApp}, we show some examples of the fits of Eq.~\ref{eqD} used to compute $D$.
    It presents values from young (left panel) and old (right panel) populations at time 13\,Gyr as seen in Fig.~\ref{fig_DandDensity}, for an arm-like grid point ($(X,Y)$\,=\,$(0,5)$\,kpc, lowest $D$) and an interarm-like one ($(X,Y)$\,=\,$(-5,0)$\,kpc, highest $D$).
    As explained in Sect.~\ref{sectMethod}, arms and interarms are more distinct in young stars than in old ones.
    
    We have chosen the contour $D_\mathrm{th}$\,=\,1.9 for the young stars (<\,0.5\,Gyr) to delimit the spiral arms regions. This choice is somewhat arbitrary because there is a continuous trend from low to high $D$ values, but is motivated by the following considerations. Since, from Fig.~\ref{fig_DandDensity}, we see that spiral arms are more prominent, well-defined and extended in radius for young stars, we take this population as reference.
    By construction, the contour of $D$\,=\,2 is approximately the one that divides the disc into two equal areas for most of the snapshots and populations (bottom panel of Fig.~\ref{fig_Dkde}). Therefore, a slightly smaller value of $D_{\mathrm{th}}$ is more restrictive in defining spiral arms. Also, contrary to old populations that have less steep arm-vs-interarm transitions, arms defined by young stars remain largely stable under small variations in the threshold (a comparison between contours 1.9 and 2 can be seen in Fig.~\ref{fig_DandDensity}) and thus, our choice gives a robust definition of the arms location.
    
%------------------------------------------
\section{Time evolution of the arm-vs-interarm contrast}\label{app_contrasts}

    \begin{figure}[h]
        \includegraphics[width=0.49\textwidth]{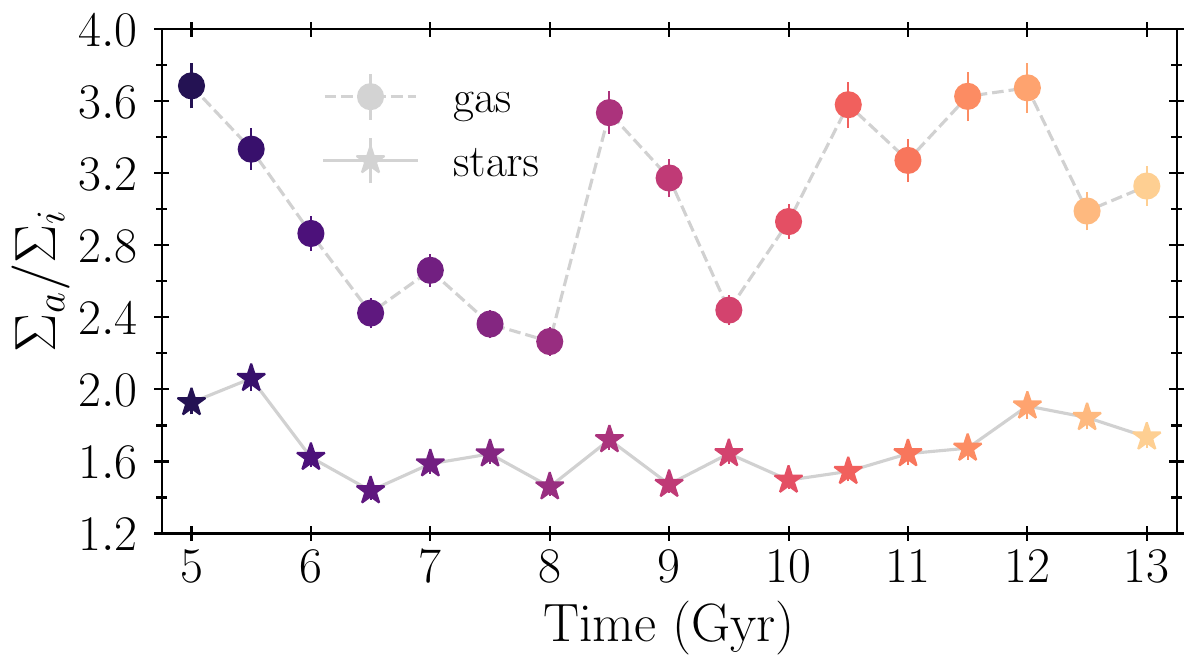}
        \caption{Same as Fig.~\ref{fig_surfdenscontrasts}, but showing arm-vs-interarm density contrasts instead of the relative overdensity of arms with respect to the disc average.}
        \label{fig_contrastsApp}
    \end{figure}
    
    Figure~\ref{fig_contrastsApp} shows the temporal evolution of the arm-vs-interarm density contrasts $\Sigma_{a}/\Sigma_{i}$ of gaseous (circles) and stellar (stars) arms. These contrasts have a low dependence on the threshold value $D_{\mathrm{th}}$ adopted. Globally, the stellar arm-vs-interarm contrast scale with $\delta_{\Sigma}$ in Fig.~\ref{fig_surfdenscontrasts} (contrast of the arms with respect to the average of the disc) and show a similar time evolution with strong arms at early and late times. 
    The gas arm-vs-interarm contrast shows a very clear peak around 8.5\,Gyr, which is not so prominent in Fig.~\ref{fig_surfdenscontrasts}. At those times, gaseous arms are about 3.6 times denser than interarms, but about 1.8 times denser than the average density of the disc. This discrepancy is explained because the interarms surface density is more prone to change than the average disc surface density.
    We note that $\Sigma_a/\Sigma_i$ can be significantly affected by the small values of the density in the interarm in the denominator, which can be especially low for the gas. The values of $\delta_{\Sigma}$ do not suffer from this and thus are more robust.

%------------------------------------------
\section{Stars traced between snapshots}\label{app_traceArms}

\begin{figure}[h]
        \includegraphics[width=0.5\textwidth]{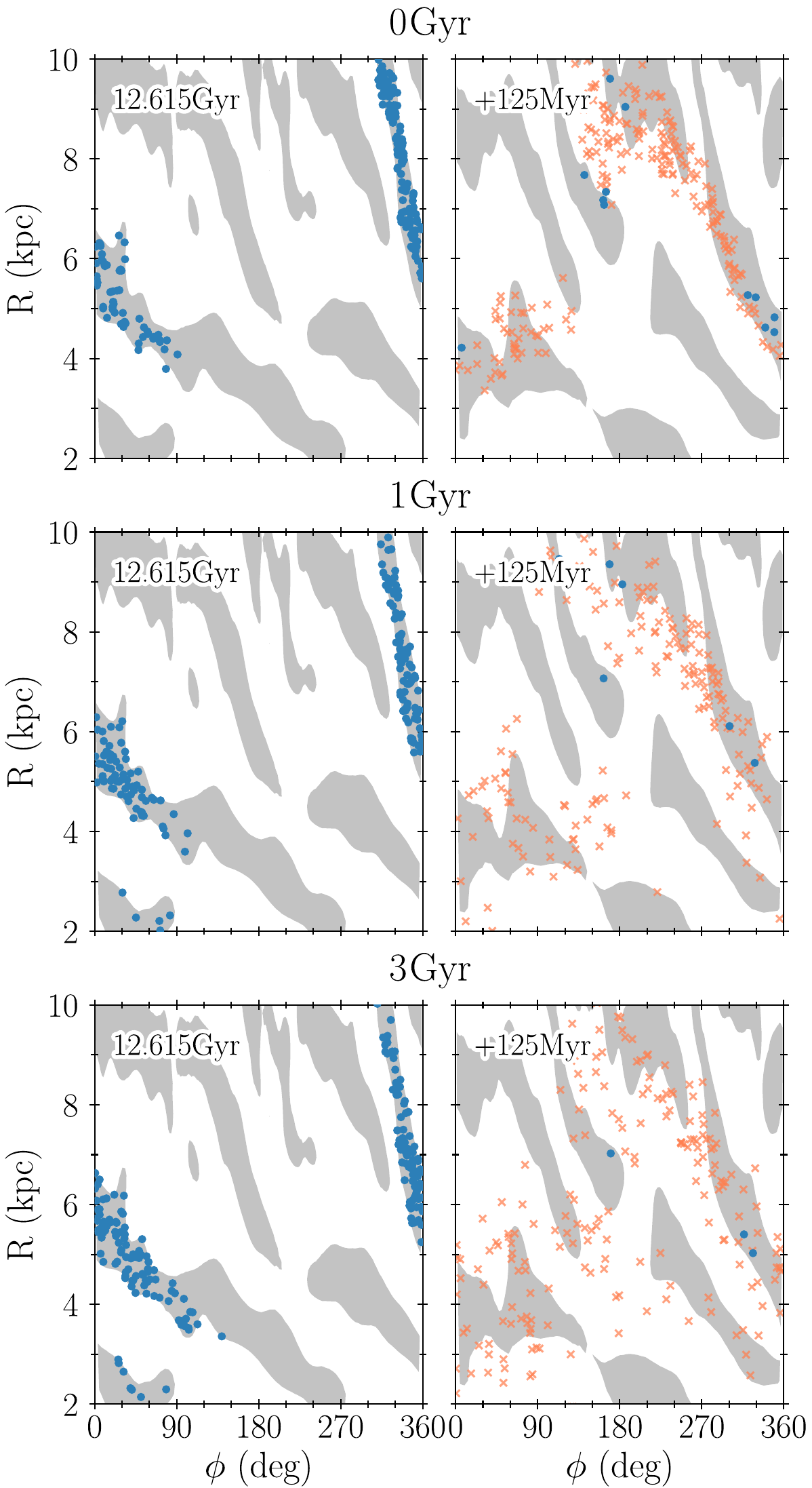}
        \caption{Same as Fig.~\ref{fig_diffusionFollowOld}, but starting from another time and including stars born within the spiral arm (first row). The starting arm in the left panels breaks and evolves into the four major arm segments at $\phi\gtrsim180^{\circ}$ in the right panels panels after slightly more than a half rotation. The rotation is in the direction of increasing azimuth}
        \label{fig_diffusionFollowApp}
    \end{figure}

    Figure~\ref{fig_diffusionFollowApp} shows stars coincident with an arm at time 12.615\,Gyr (left column) and their location 125\,Myr later (right column). In the left panels, stars have ages smaller than 5\,Myr (top), from 1 to 1.005\,Gyr (centre) and from 3 to 3.005\,Gyr (bottom). This figure clearly shows the change in behaviour of different stellar populations selected in the same arm described in Sect.~\ref{sectTrace}.
    
%------------------------------------------
\section{Torques maps}\label{app_torquesmaps}

\begin{figure}
        \begin{center}
        \includegraphics[width=0.3\textwidth]{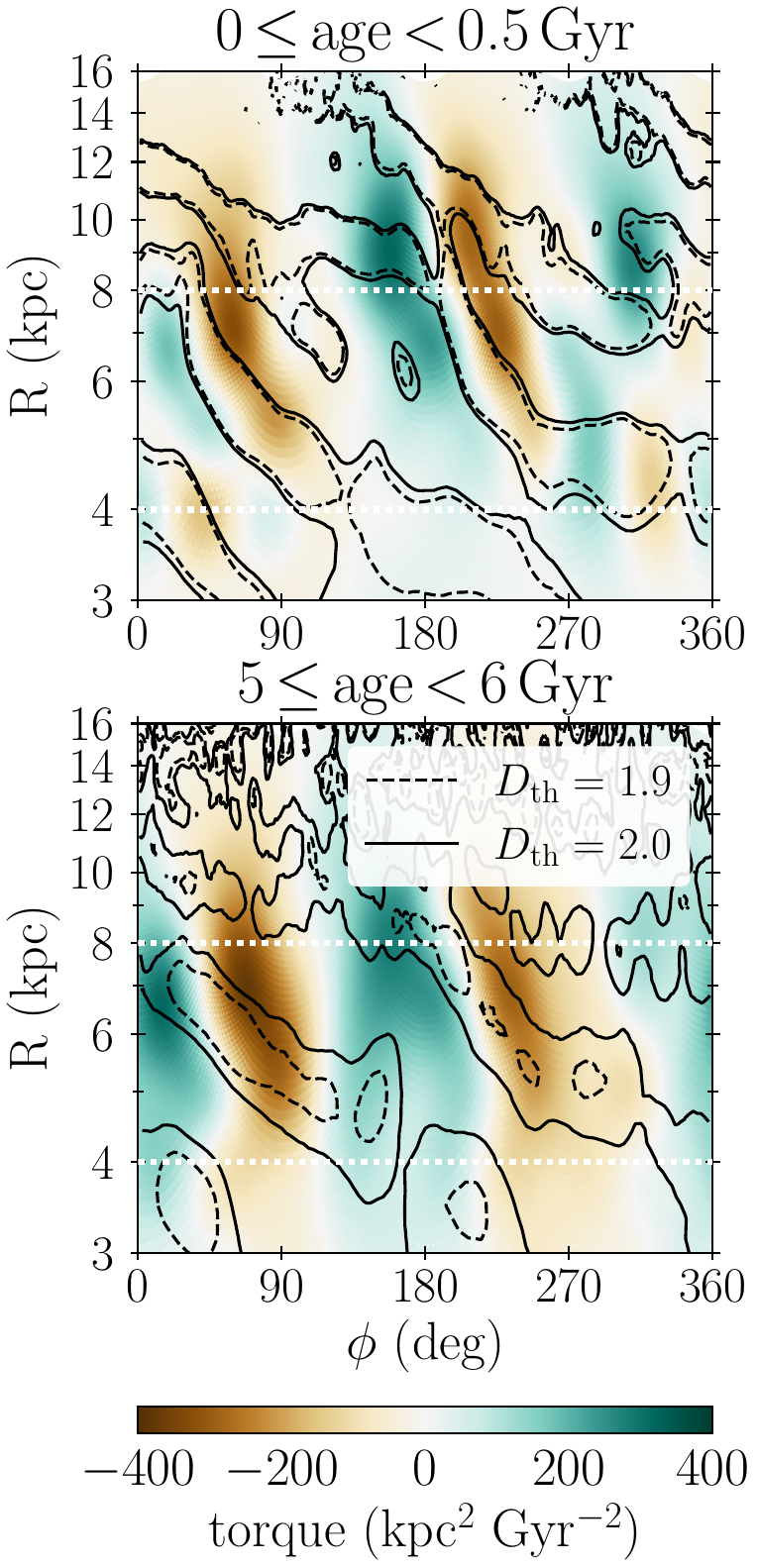}
        \caption{Maps of torques created by stars younger than 0.5\,Gyr (top) and by stars with ages between 5 and 6\,Gyr (bottom) at time 13\,Gyr.
        Stars inside 4\,kpc were azimuthally shuffled before computing the torques.}
        \label{fig_torquesXY}
        \end{center}
    \end{figure}

    Figure~\ref{fig_torquesXY} shows an example of the in-plane map of torques created by two stellar populations at time 13\,Gyr. In both cases, we avoided bar torques by azimuthally shuffling stars with galactocentric radii smaller than 4\,kpc.
    
\end{appendix}

%%%%%%%%%%%%%%%%%%%%%%%%%%%%%%%%%%%%%%%%

\end{document}